\documentclass[%
 reprint,
nofootinbib,
 amsmath,amssymb,
 aps,
floatfix,
]{revtex4-2}

\usepackage{multirow}
\usepackage{graphicx}
\usepackage{dcolumn}
\usepackage{bm}
\usepackage{hyperref}
\usepackage{dsfont}
\usepackage{subfigure}
\usepackage{comment}
\usepackage{color}
\usepackage[normalem]{ulem}
\usepackage{soul}
\usepackage{float}
\usepackage{subfiles}

\begin{document}

\preprint{APS/123-QED}

\title{Detecting the heterodyning of gravitational waves}
\author{Jakob Stegmann$^{1,3}$}
 \email{jstegmann@mpa-garching.mpg.de}
\author{Sander M. Vermeulen$^{2,3}$}%
 \email{smv@caltech.edu}
\affiliation{
 $^{1}$Max Planck Institute for Astrophysics, Karl-Schwarzschild-Straße 1, 85741 Garching b. München, Germany\\
 $^{2}$California Institute of Technology, Department of Physics, Pasadena, California 91125, USA\\
 $^{3}$Gravity Exploration Institute, School of Physics and Astronomy, Cardiff University, Cardiff, CF24 3AA, UK
}


\date{\today}

\begin{abstract}
Gravitational waves modulate the apparent frequencies of other periodic signals. Low-frequency gravitational waves could therefore be detected by observing frequency modulations in signals from higher-frequency sources, e.g., those from binary white dwarfs detected with the LISA gravitational-wave detector. We propose a concrete method to extract these modulations by coherently adding the cross-spectra of a large number of well-resolved quasi-monochromatic signals. We apply this method to the case of LISA, and find this method would enable the detection of background gravitational wave strain amplitudes of, e.g., $A\simeq10^{-10}$ at a frequency $F\simeq10^{-8}\,\rm Hz$, given current projections for the number and properties of Galactic binary white dwarfs and the sensitivity of the instrument. We also estimate (to within an order of magnitude) that this method could potentially compete with that of current Pulsar Timing Arrays when using signals from binary neutron stars such as those expected to be observed with proposed detectors like DECIGO. Our results show that gravitational-wave detectors could be sensitive at frequencies outside of their designed bandwidth using the same infrastructure, which has the potential to open up unexplored and otherwise inaccessible parts of the gravitational wave spectrum.
\end{abstract}

\maketitle
\section{Introduction}
The field of gravitational-wave astronomy, as established with the first direct detection of gravitational waves (GWs) \citep{GW150914}, is still in its infancy. So far, only GWs with frequencies between $\sim10-500$~Hz produced by the coalescence of black holes and neutron stars with masses $\sim1-100$ times the mass of our Sun have been detected \citep{2021arXiv211103634T}. New detectors and techniques are being developed to probe different regions of the GW frequency spectrum and to investigate numerous other potential GW sources; e.g., rotating neutron stars \citep{1998PhRvD..58h4020O}, binary white dwarfs (BWDs) \citep{2001A&A...375..890N}, intermediate-mass and super-massive binary black holes (SMBHBs) \citep{2016PhRvD..93b4003K}, a background of primordial GWs \citep{2006PhRvD..73f3008M}, and dark matter \cite{2016PhRvD..94h3504C,brito_gravitational_2017}. 

The sensitive bandwidth of laser interferometers (the only proven type of GW detector), is typically limited at low frequencies by spurious accelerations of the test masses, and at high frequencies by quantum uncertainty in the optical state and an intrinsically decreased response to GWs with wavelengths shorter than the interferometer's arms. Laser interferometers can be very sensitive at higher frequencies ($\sim1-100$~MHz), using 
cross-correlation and shorter arms \citep{vermeulen_experiment_2021,chou_holometer:_2017}. Increasing the sensitivity at lower frequencies is not straightforward, and even a space-based instrument such as LISA \citep{2017arXiv170200786A}, subject to greatly reduced environmental noise, will not be sensitive to GWs below $\lesssim10^{-5}$~Hz. While marginal gains have been made in understanding and addressing the complex amalgam of low-frequency noise contributions encountered in laser interferometers (which include fundamental quantum limits) \citep{buikema_sensitivity_2020}, it seems unlikely that their bandwidth will expand into lower frequencies by more than an order of magnitude in the coming decades.  

Other detection techniques to probe new areas of the GW spectrum have been proposed and some have been tried. At high frequencies ($\rm kHz$~--~$\rm GHz$) these include techniques that exploit graviton-to-photon conversion (known as the inverse Gertsenshtein effect) \citep{gertsenshtein_wave_1962,ejlli_upper_2019}, optically levitated sensors, resonant mass detectors \citep{goryachev_gravitational_2014}, and more \citep{aggarwal_challenges_2021}. 

At low frequencies, currently the only competitive method to search for GWs is using sets of time-resolved observations of pulsars, known as Pulsar Timing Arrays (PTAs), which are sensitive in the $\rm nHz$~--~$\mu\rm Hz$ range  \citep{1979ApJ...234.1100D,1978SvA....22...36S,1982MNRAS.199..659M,1983MNRAS.203..945B,1983ApJ...265L..39H,1990ApJ...361..300F,1994ApJ...428..713K,2005ApJ...625L.123J,2006ApJ...653.1571J,2009MNRAS.394.1945H,2010MNRAS.407..669Y,2016MNRAS.458.1267V,2016MNRAS.455.1665B,2020ApJ...905L..34A}. GWs in this frequency range provide a probe of the most massive binary black holes ($\gtrsim10^9\,\rm M_\odot$) in the Universe, which are expected to form at the centre of merging galaxies \citep{2005SSRv..116..523F}. Measuring the low-frequency GWs from individual SMBHBs or a stochastic background thereof allows investigation of their abundance and coevolution with their host galaxies. Several PTAs across the globe have recently found the first evidence for a low-frequency GW background \citep{2023ApJ...951L...8A,2023A&A...678A..50E,2023ApJ...951L...6R,Xu_2023}, by monitoring the arrival times from dozens pulsars for more than a decade. GWs incident on a pulsar and/or the detector produce deviations of the apparent frequency or equivalently the arrival time of the radio pulses that are correlated between different pulsars. This detection technique thus exploits the interplay of electromagnetic pulses with GWs which results in a modulation of the pulse frequency. 

Recently, work by \citet{2022PhRvD.105d4005B} was published that proposed a new method to look for low-frequency gravitational waves\footnote{We note that our work was developed independently from~\citet{2022PhRvD.105d4005B}, and we became aware of their work only after drafting this manuscript.}. This method for detecting (low-frequency) GWs uses the interactions between GWs of different frequencies. The basis of the method is the gravitational red- and blueshift induced by one GW onto the other. This mechanism can also be viewed as one GW perturbing the space-time along the direction of travel of the other GW, and thus modulating the arrival times of peaks and troughs of the other GW. Mathematically, the effect can be described as a multiplication or mixing of two GWs. From this description, it can be shown that the resulting GW signal contains Fourier components at the sum and difference of the frequencies of the two waves, with an amplitude proportional to the product of the amplitudes of the individual GWs. This elementary result of the mixing of two waves, also known as heterodyning, has been used in the processing of electromagnetic signals for over a century. Heterodyning effectively produces a frequency-shifted copy of one signal (known as a sideband) in the frequency range of a readily detectable second signal. As we show in this paper, this mechanism can be used in GW astronomy, where GW signals detectable with, e.g., laser interferometers can be used to detect low-frequency background GWs. This method of searching for low-frequency GWs is conceptually similar to the technique used by PTAs, with the crucial difference that instead of looking for disturbances in the periodic electromagnetic signal of pulsars, we look for disturbances in a periodic GW signal. 

This method allows one to expand the sensitive bandwidth of GW detectors into low-frequency regimes using the detectors' existing infrastructures. Moreover, this method could enable a sensitivity to GWs in a bandwidth where no other detection methods exist, e.g., in the $\mu\mathrm{Hz}$~regime where the frequency ranges of space-based laser interferometers and PTAs leave a gap.

The work presented here is complementary to the work by~\citet{2022PhRvD.105d4005B}; we demonstrate a concrete cross-correlation search method for detector data and we simulate signals expected to be detected with LISA using an observationally-driven mock population of BWDs. In addition, as the heterodyning method is applicable to general periodic GW signals, we present order-of-magnitude estimates of the sensitivity that could be obtained with DECIGO \citep{DECIGO}, ET, and CE, which are expected to be able to observe large numbers of GW signals from BWDs and binary neutron stars (BNSs). The analysis and projections of the potential of the method in \cite{2022PhRvD.105d4005B} for observations with LISA differ from ours; in Sec.~\ref{sec:discussion} we expand on the difference in findings. 

\section{Theory}\label{sec:Single-SMBHB}
 \begin{figure}
 \includegraphics[width=1.\columnwidth]{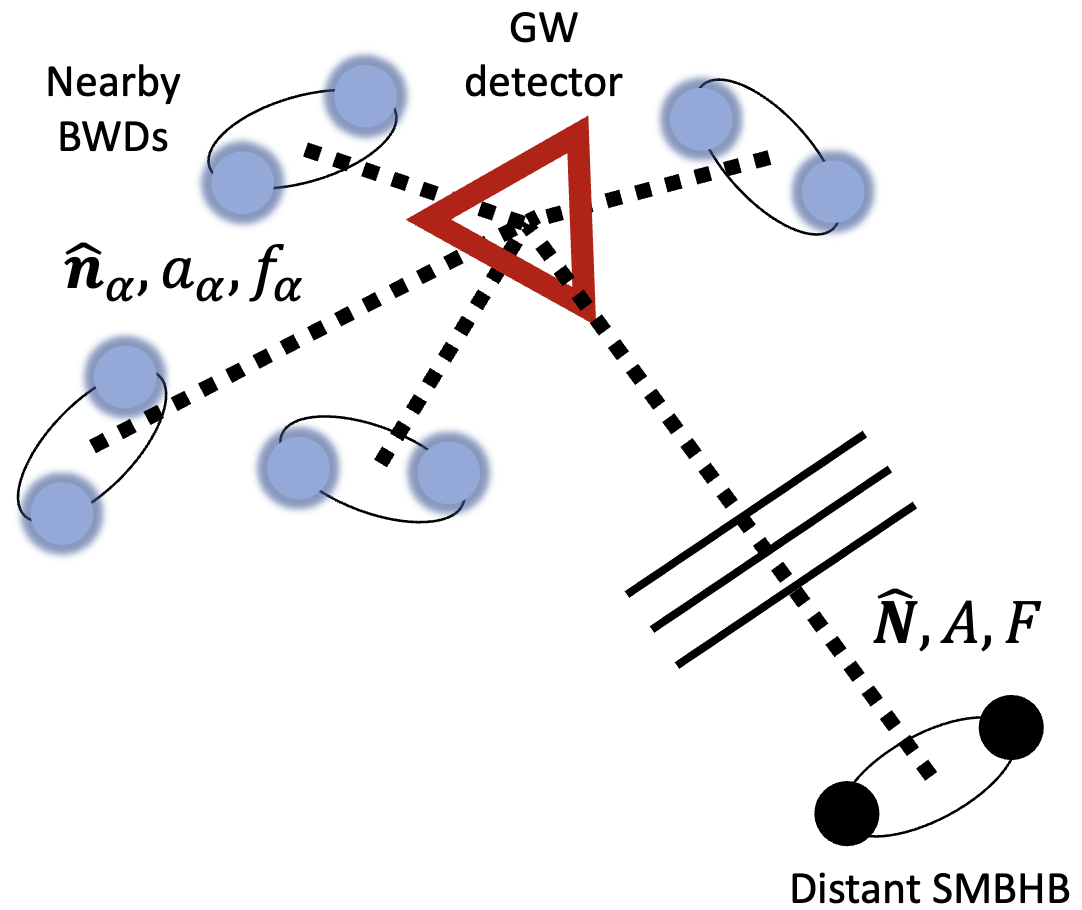}
 \caption{Sketch of the GW heterodyning method proposed in this work. The LISA GW detector aims to resolve GW signals from a large number of BWDs in the Milky Way \citep{2017arXiv170200786A}. If there is another GW, e.g., one emitted by a much more distant SMBHB, this could be indirectly detectable through the coherent modulations imparted on the measured GW frequency of the BWDs. Figure inspired by \citet{2022PhRvD.105d4005B}.}
 \label{fig:sketch}
\end{figure} 
We consider a set of $N\gg1$ periodic GW sources which could be simultaneously observed for a long time (e.g., BWDs in our Galaxy that could be individually resolved by LISA \citep{2017arXiv170200786A}). We further assume that these sources emit quasi-monochromatic GWs, i.e., that their frequency does not significantly change within the observation time $T$ (see Sec.~\ref{sec:discussion} for discussion of the implications of relaxing this assumption). In that case we can write the GW signal (in units of strain) from the $\alpha$-th periodic source at distance $d_\alpha$ as 
\begin{align}\label{eq:monochromatic}
    &h_\alpha(t)=a_\alpha\cos[2\pi f_\alpha t+\varphi_\alpha],&(\alpha=1,2,\dots,N),
\end{align}
with constant frequency $f_\alpha$, amplitude $a_\alpha$, and initial phase $\varphi_\alpha$. We refer to these GWs as carrier signals and to their sources as carrier sources.

If there is an incident GW from a secondary, more distant source, this GW will perturb the spacetime at the location of the carrier sources and at the location of the observer (see Figure~\ref{fig:sketch}). As a consequence, the frequency of the GW carrier signals are no longer constant but are modulated in time. For a background GW emitted by a distant point source in the direction $\bm{\hat{N}}$ this frequency modulation of the carrier signal is given by \citep{MaggioreMichele2018GWV2}
\begin{align}\label{eq:modulation}
    \frac{f_\alpha-f_\alpha(t)}{f_\alpha}=&\frac{n^i_\alpha n^j_\alpha}{2(1+\bm{\hat{N}}\cdot\bm{\hat{n}}_\alpha)}\left[h_{ij}^{\rm TT}(t)-h_{ij}^{\rm TT}(t_{\alpha})\right],
\end{align}
where $\bm{\hat{n}}_\alpha$ and $n^i_\alpha$ is the unit vector from the observer to the $\alpha$-th carrier source and its components, respectively, and $t_\alpha = t-d_\alpha(1+\bm{\hat{N}}\cdot\bm{\hat{n}}_\alpha)/c$ is the retarded time coordinate that accounts for the propagation of the carrier wave. We will refer to the lower-frequency GW of interest that modulates the carrier signal as the background GW. Additionally, $h_{ij}^{\rm TT}(t)$ and $h_{ij}^{\rm TT}(t_{\alpha})$ correspond to the metric perturbation due to the incident GW at the spacetime locations of the carrier source and the observer, respectively (in the terminology of PTAs \citep{1979ApJ...234.1100D,1978SvA....22...36S}, the former is usually referred to as the `Earth term' and the latter as the `pulsar term').

It can be shown that the single-sided frequency spectrum of the modulated signal can then be written as \citep[][]{2022PhRvD.105d4005B}
\begin{align}\label{eq:modulated-spectrum}
     \tilde{h}_{\alpha}(f)&\simeq a_{\alpha} e^{i \varphi_{\alpha}} \delta (f_{\alpha}-f)\nonumber\\
     &+\frac{1}{2} a_{\alpha}A I_{\alpha,L} e^{i ( \varphi_{\alpha}+ \Phi_L)} \delta (f-f_{\alpha}+F_L)\nonumber\\
     &+\frac{1}{2} a_{\alpha}A I_{\alpha,L} e^{-i (\varphi_{\alpha}+ \Phi_L)} \delta (f-f_{\alpha}-F_L)\nonumber\\
     &+\frac{1}{2} a_{\alpha}A I_{\alpha,D} e^{i (\varphi_{\alpha}+ \Phi_{\alpha,D})} \delta (f-f_{\alpha}+F_{D,\alpha})\nonumber\\
     &+\frac{1}{2} a_{\alpha}A I_{\alpha,D} e^{-i ( \varphi_{\alpha}+ \Phi_{\alpha,D})} \delta (f-f_{\alpha}-F_{D,\alpha}),
\end{align}
where $I_{\alpha,L,D}=\left({f_\alpha}/F_{L,D}\right)\, K(\bm{\hat{N}},\bm{\hat{n}}_\alpha,h_{ij}^\mathrm{TT},d_\alpha)$, and $K$ is a purely geometrical factor of order unity that accounts for the polarisation, propagation direction, and propagation distance of the background and carrier GWs. The first term in the spectrum given by Eq.~\eqref{eq:modulated-spectrum} is the Fourier component corresponding to the carrier signal at the frequency $f=f_{\alpha}$. The modulation due to the background GW at the location of the observer manifests as two Fourier components with frequencies  $f=f_{\alpha}\pm F_L$ (second and third term in Eq.~\ref{eq:modulated-spectrum}), which we will refer to as the `local' sideband terms. Similarly, the modulation of the carrier signal due to the background GW at the location of the carrier source produces sidebands with frequencies $f=f_{\alpha}\pm F_{D,\alpha}$ (fourth and fifth term), which we will refer to as the `distant' sideband terms. Note that the frequency $F_L$ and phase offsets $\Phi_L$ of the `local' terms are independent of the carrier (they are equal to the frequency and phase of the modulating GW at the location of the observer), whereas the `distant' terms have frequency and phase offsets $F_{D,\alpha}$ and $\Phi_{\alpha,D}$, which depend on the location of the carrier source. 

This mechanism, a sort of `GW heterodyning' could allow the indirect detection of low-frequency GWs that may otherwise be undetectable when a GW detector is not sensitive to signals down to a frequency $F$, but is sensitive at much higher frequencies $f_\alpha+ F$. Using this method, the upconverted background signal amplitude is $A_\mathrm{sideband}=Aa_{\alpha}Kf_\alpha/F$. For example, if we take the carrier signal to be the GWs emitted by a typical BWD with amplitude $a_\alpha$ (such as the BWDs that LISA aims to detect), and with frequency $f_{\alpha}\sim 10^{-2}\,\rm Hz$, and we take the background signal to be GWs emitted by a SMBHB with amplitude $A\sim10^{-12}$ and frequency $F_L\sim 10^{-8}\,\rm Hz$, the background sideband signal appears at an amplitude $A_\mathrm{sideband} \sim 10^{-6}a_\alpha$.

This suppression relative to the carrier would mean the background signal amplitude is below the typical noise level of the detector. In the following section, we propose a method to amplify the signal which utilises the coherence of the modulation of multiple carrier signals. To this end, we construct and add $N_p=N(N-1)/2\gg1$ different cross-spectra (one for each pair of carrier sources) such that the sideband terms sum up coherently to exceed the incoherent random noise.

\section{Cross-correlation analysis}\label{sec:data-analysis}
We propose a cross-correlation method for detecting a background gravitational wave signal that produces phase modulation of carrier GW signals. We will later use this method to make quantitative estimates of the expected signal-to-noise ratio that can be obtained for potential astrophysical GW sources using planned GW detectors. 

We consider the time-domain output signal of the GW detector $s(t)$ to be given by the sum of $N$ carriers, all modulated by a single background GW signal with frequency $F$ corresponding to either the `local' ($F=F_L$) or the `distant' ($F=F_D$) term, and noise $n(t)$ characteristic of the detector:
 \begin{equation}
     s(t) = \sum_{\alpha=1}^N h_\alpha(t) + n(t).
 \end{equation}
For any carrier, we can apply a demodulation and phase-shift to the time-domain detector output and normalise it by the modulation index and the carrier amplitude:
\begin{align}\label{eq:demodulation}
     &s_\alpha(t)= \frac{\sqrt{2}}{a_\alpha I_\alpha} e^{-i (2\pi f_\alpha t + \varphi_\alpha)}\; s(t).
\end{align}
This demodulation shifts the frequency of all Fourier components in the output by an amount $f_\alpha$, such that all sideband (heterodyne) signals are frequency shifted to the frequency $\pm F$ of the modulating background GW that produces them. Moreover, any heterodyne signals from background GWs will now appear with a Fourier amplitude equal to the background GW strain amplitude that produces them. In general, the demodulation frequency need not be constant in time, but could be adjusted over time to account for time-dependent changes in the carrier frequency. Specifically, the demodulation frequency and phase could be varied according to a predetermined carrier signal model, or they could be fit to the data post hoc (e.g., through maximising the demodulated carrier amplitude) when the frequency evolution is unknown a priori. After this frequency and phase shift, we can apply an appropriate low-pass filter to the data such that other terms, as long as they are well-separated from the carrier and modulation sideband, need not be considered\footnote{The sidebands due to the local modulation can be considered well-separated in the frequency domain from the sidebands due to the distant modulation (pulsar/distant term) when ${|F_L-F_D|\gg 1/T}$. We also assume all carrier signals are well-separated from each other (${|f_\alpha-f_\beta| \gg 1/T \; \forall\; \alpha,\beta}$).}.  

We consider the case where the time-domain detector output is discretised with a constant sampling frequency $f_s$ for a total observation time $T$. Next, we take the single-sided discrete Fourier transform of the detector output, which yields a discrete complex amplitude spectrum $S^j_\alpha$ for each carrier signal, which will have the form 
\begin{align}\label{eq:carrierspectrum}
   &S^j_\alpha= A e^{{i}\Phi_\alpha}\delta^j_{l(F)} +  \frac{\sqrt{2}}{a_\alpha I_\alpha}\sqrt{\frac{\rho^j_\alpha}{T}}\;e^{i\eta^j_\alpha},
\end{align} 
where the index $j=1,2,\dots,Tf_s/2$ runs over the frequency bins, $l(F)$\footnote{$l(F)=\lceil FT +\frac{1}{2} \rceil$} is the index of the bin that contains the background signal ($\delta^j_{l}$ is the Kronecker delta), $\rho^j_\alpha$ is the noise power spectral density of the detector, and $\eta^j_\alpha$ are the random noise phases (where both noise parameters have undergone the frequency and phase shift described by Eq.~\ref{eq:demodulation}). 
The spectrum $S^j_\alpha$ is unique for each carrier signal. As background GWs would modulate all carrier signals coherently (i.e., the sideband phase is deterministic), whereas the noise has a random phase, cross-correlating different carrier signals is advantageous. For each pair of carrier signals $(\alpha\beta)$, a cross-spectrum $S^j_{\alpha\beta} =  S^{j}_\alpha S^{j*}_\beta$ can be constructed which has the form
\begin{equation}
     S^j_{\alpha\beta}= A^2 e^{{i}(\Phi_\alpha - \Phi_\beta)}\delta^j_{l(F)} + \frac{2}{a_\alpha a_\beta I_\alpha I_\beta}\frac{\sqrt{\rho^j_\alpha\rho^{j}_\beta}}{T} \;e^{i\left(\eta^j_\alpha- \eta^j_\beta\right)},
\end{equation}
where $\Phi_\alpha - \Phi_\beta=\Phi_{\alpha \beta}$ is the phase difference of the modulating signal between the two carrier signals. From this expression it can be seen that $\Phi_{ab}$ is deterministic, and $\eta^j_\alpha- \eta^j_\beta=\eta^j_{\alpha\beta}$ is random. Therefore, we can add up signal terms from different cross-spectra coherently, and the noise will average out. If we have $N$ individually resolved carriers at our disposal we can construct $N_p=N(N-1)/2$ different cross spectra and take a coherent weighted average of them
\begin{equation}\label{eq:avg_cross_spectrum}
    S^j = \frac{\sum_{(\alpha\beta)}^{N_p} \lambda^j_{\alpha\beta} S^j_{\alpha\beta}\, e^{-i\Phi_{\alpha\beta}}}{\sum^{N_p}_{(\alpha\beta)} \lambda^j_{\alpha\beta}}, 
\end{equation}
where $\lambda^j_{\alpha\beta}$ are the weights of each cross-spectrum. Performing this coherent summation is possible as long as the relative modulation sideband phase $\Phi_{\alpha\beta}$ can be determined for each carrier pair ($\alpha\beta$). For the modulation produced by the background GW at the detector (`local' term), $\Phi_{\alpha\beta}=0 \;\forall\, \alpha\beta$. For the sideband due to the modulation produced at the source of the carrier GW signal (`distant' term), $\Phi_{\alpha\beta}$ is a function of the relative positions of the background GW source and the carrier signal sources. In this case, $\Phi_{ab}$ can be taken as free parameters that are fit to the data by maximising the total SNR for a particular sideband frequency, which would yield an upper estimate of the maximum background GW signal power at a certain frequency. Alternatively, a hypothetical background source position and frequency could be assumed, which prescribes a certain set of $\Phi_{\alpha\beta}$ given the geometry of the source positions, which would then yield an upper limit of the estimated background GW strain at that frequency and sky position. 

Note that the coherent average is constructed such that the expected real part of the signal bin is ${\mathrm{E}\left[\mathrm{Re}[S^{l(F)}]\right]=A^2}$. The squared signal-to-noise ratio can thus be defined for each bin
\begin{equation}
(\mathrm{SNR}^j)^2 = \frac{\left(\mathrm{Re}[S^j]\right)^2}{\mathrm{Var}\left(\mathrm{Re}[S^j] \right)}.
\end{equation}
It can be shown that an optimal signal-to-noise ratio is found by taking the weights \citep{PhysRevD.103.063027}
\begin{equation}
    \lambda^j_{\alpha\beta} = \sum_{(\gamma\delta)}^{N_p} ([C^j]^{-1})_{\alpha\beta,\gamma\delta}\simeq\left(\frac{1}{\sigma^j_\alpha\sigma^j_\beta}\right)^2 = \frac{(a_\alpha a_\beta I_\alpha I_\beta)^2 T^2}{4\rho^j_\alpha \rho^j_\beta},
\end{equation}
where $C^j_{\alpha\beta,\delta\gamma}$ is the pair-wise cross-covariance matrix of the cross-spectra $S_{\alpha\beta}^j,S_{\delta\gamma}^j$, and $\sigma^j_{\alpha,\beta}$ are the variances of frequency bin $j$ in each carrier spectrum (Eq.~\ref{eq:carrierspectrum}); the approximation holds in the weak-signal limit \citep{PhysRevD.103.063027}. The SNR of a modulating background GW with frequency $F$ and amplitude $A$ can now be evaluated:
 \begin{equation}
      (\mathrm{SNR}^{l(F)})^2\simeq \frac{ A^4}{2}\mathcal\sum_{(\alpha\beta)}^{N_p}\left(\frac{1}{\sigma^{l(F)}_\alpha\sigma^{l(F)}_\beta}\right)^2\label{eq:snr}.
 \end{equation}

 \section{Methods}
The GW detector LISA is expected to observe a large number of continuous, periodic GW signals from BWDs in our Galaxy \citep{1990ApJ...360...75H,2001A&A...375..890N,2009ApJ...693..383R,2010ApJ...717.1006R,2012A&A...546A..70T,2017arXiv170200786A,2019MNRAS.490.5888L,2022MNRAS.511.5936K}. These BWDs could potentially serve as carrier sources that allow for the detection of low-frequency background GWs as described above. 

The total number and properties of Galactic BWDs is subject to large uncertainty. To obtain a quantitative projection for the number, frequency, and amplitude of BWD GW signals that may be detected with LISA, we use an observationally driven parametric model of the Galactic white dwarf population, constructed by \citet{2022MNRAS.511.5936K}\footnote{\url{https://gitlab.in2p3.fr/korol/observationally-driven-population-of-galactic-binaries}, accessed January 6, 2023.}. This model builds upon the spectroscopic samples of single white dwarfs and BWDs from the Sloan Digital Sky Survey (SDSS) and the Supernova Ia Progenitor surveY (SPY) to produce a synthetic population of Galactic BWDs which are specified by their component masses, orbital frequencies, sky positions, and orientations. These source parameters are then used to calculate the GW signals of each BWD in the population. Part of the BWDs would emit GWs at low frequencies $f\lesssim3\,\rm mHz$ and are predicted to be so numerous that they are not individually resolvable but constitute a confusion-limited foreground noise \citep{2010ApJ...717.1006R}. The rest, an estimated number of $\sim\mathcal{O}(10^3\,$--$\,10^5$) BWDs emit GWs at higher frequencies and are expected to be sufficiently loud that they are individually resolvable; these are the BWDs which can be used as carrier sources in our method. 

\begin{table}
	\centering
	\caption{Input parameters used for generating synthetic populations of Galactic binary white dwarfs. The parameters $\rho^{\rm Korol}_{\rm WD,\odot}$, $f^{\rm Korol}_{\rm BWD, 4\, AU}$, $f^{\rm Korol}_{{\rm BWD}, a_{\rm max}}$, and $\alpha^{\rm Korol}$ are used as input for the algorithm described by \citet{2022MNRAS.511.5936K} to model the sets of BWD carrier signals. These parameters represent the local WD density, the fraction of binaries with semi-major axes~$<4\,\rm AU$, the fraction of binaries with semi-major axes less than the maximum separation detectable with LISA ($a_{\rm max}$), and a power-law index specifying the BWD semi-major axis distribution, respectively (see \citet{2022MNRAS.511.5936K} for details). The values of these parameters were chosen to correspond to upper ({\tt{Optimistic}}), median ({\tt{Moderate}}), and lower ({\tt{Pessimistic}}) observational limits. We chose observation times $T$ between $1.0$ and $10.0\,\rm yr$. $N$ indicates the resulting number of BWDs which are individually resolvable with LISA.}	
	\label{tab:parameters}
    \begingroup
	\renewcommand*{\arraystretch}{1.2}
	\begin{tabular}{ccccc}
	\hline
	\hline
    \multicolumn{2}{c}{Model} & {\tt Pessimistic} & {\tt Moderate} & {\tt Optimistic} \\
    \hline
	\hline
    {$\rho^{\rm Korol}_{\rm WD,\odot}$} & $[10^{-3}\,{\rm pc}^{-3}]$ & $4.11$ & $4.49$ & $4.87$\\
    $f^{\rm Korol}_{\rm BWD, 4\, AU}$ & & $0.112$ & $0.095$ & $0.078$\\
    {$f^{\rm Korol}_{{\rm BWD}, a_{\rm max}}$} & & $0.008$ & $0.009$ & $0.010$\\
    $\alpha^{\rm Korol}$ & & $-1.18$ & $-1.30$ & $-1.45$\\
    $T$ & $[\rm yr]$ & $1.0$ & $4.0$ & $10.0$\\
    $N$ & & $7.0\times10^4$ & $1.1\times10^5$ & $1.9\times10^5$\\
    \hline
	\hline
	\end{tabular}
	\endgroup
\end{table}
We consider three models with different carrier source and observation parameters, {\tt Pessimistic}, {\tt Moderate}, and {\tt Optimistic}. For these models, we synthesised three BWD populations using different input parameters for the model of \citet{2022MNRAS.511.5936K}; specifically we vary the local WD density $\rho^{\mathrm{Korol}}_{\rm WD,\odot}$, the WD binary fraction $f^{\mathrm{Korol}}_{\rm BWD}$, and the power-law index $\alpha^{\mathrm{Korol}}$, which describes the BWD semi-major axis distribution (see \citet{2022MNRAS.511.5936K}). On the observation side we use three different values for the LISA mission lifetime $T=1.0$, $4.0$, and $10.0\,\rm yr$, which sets the length of observation. To get an upper and lower limit for the resulting sensitivity to background GWs, we choose the model parameters such that $\tt Pessimistic$ and $\tt Optimistic$ models yield the lowest and highest number of BWDs within the current observational uncertainty while $\tt Moderate$ model corresponds to median values. The parameter values of the three different models are summarised in Table~\ref{tab:parameters}. 

\begin{figure}
 \includegraphics[width=1.\columnwidth]{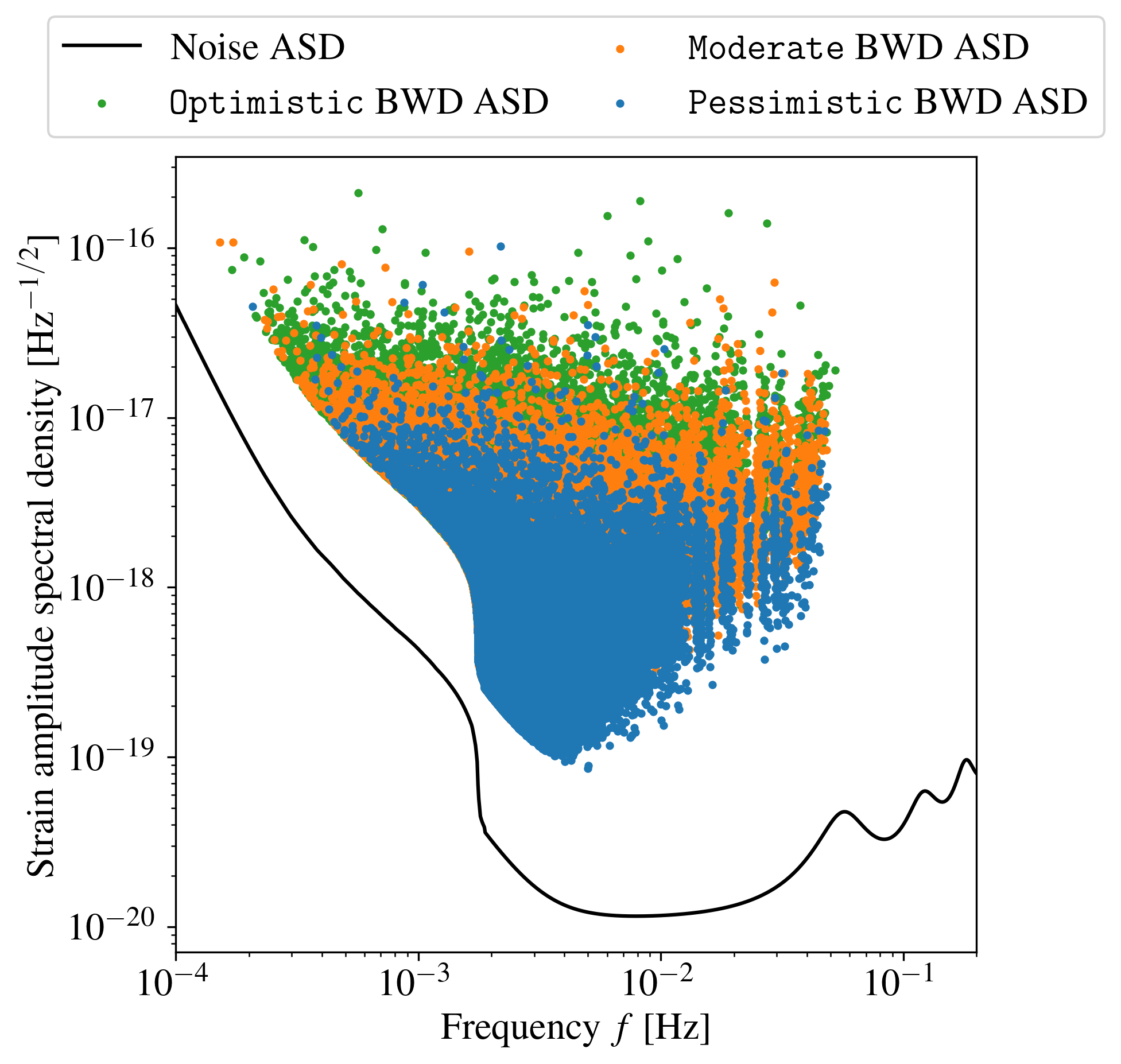}
 \caption{Amplitude spectral densities $a_\alpha\sqrt{T}$ of gravitational wave signals from individually resolvable binary white dwarfs (BWDs) in three different models \cite{2022MNRAS.511.5936K} as a function of their frequency $f=f_\alpha$. The solid line indicates the root of the projected noise power spectral density $\sqrt{\rho}$ of LISA \citep[][]{2019CQGra..36j5011R,2022MNRAS.511.5936K}. BWDs are assumed to be individually resolvable if $a_\alpha\sqrt{T/\rho(f_\alpha)}>7$.}
 \label{fig:resolvable}
\end{figure} 
In Figure~\ref{fig:resolvable}, we show the amplitude spectral density (ASD) of the BWD carriers for each model together with LISA's projected detector noise amplitude spectral density, as in \citep[][]{2019CQGra..36j5011R}, modified to account for the confusion noise due to unresolved BWDs derived by \citet{2022MNRAS.511.5936K}. Throughout this work we assume a BWD to be individually resolvable if $a_\alpha\sqrt{T/\rho(f_\alpha)}>7$, (where $\rho(f_\alpha)$ is the noise power spectral density evaluated at the BWD frequency), although the precise threshold does not affect the resulting sensitivity due to the dominant contribution of loud sources (see Section~\ref{sec:discussion}).

\section{Results}
\begin{figure*}
 \includegraphics[width=2.\columnwidth]{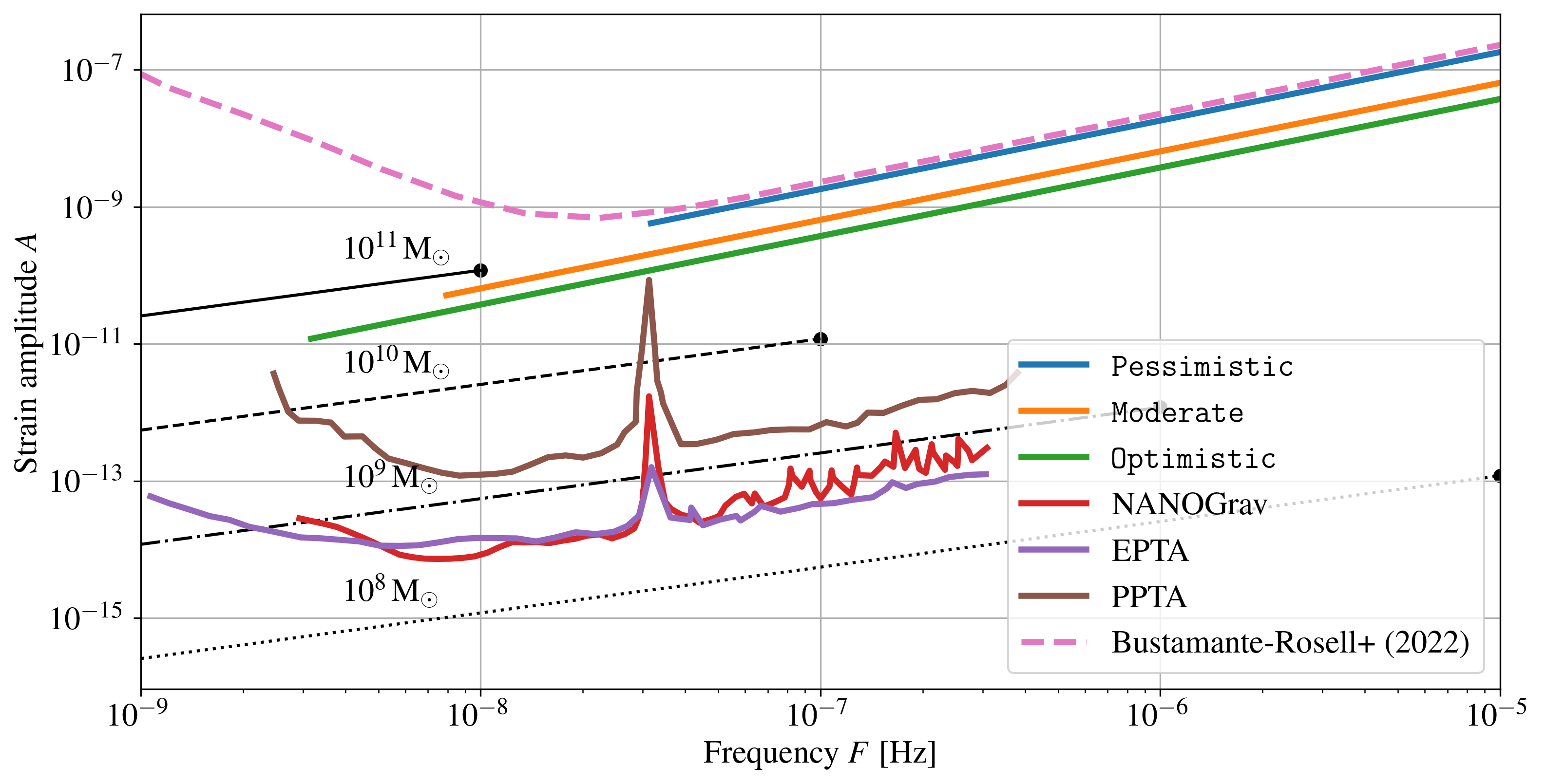}
 \caption{Sensitivity to low-frequency gravitational waves (GWs) that can be obtained by searching for correlated modulations in a set of well-resolved GW signals from binary white dwarfs (BWDs), as expected to be detected with LISA. For reference, we show the expected GW amplitudes of SMBHBs with chirp masses ranging from $10^{8}$ to $10^{11}\,\rm M_\odot$ at a fiducial distance $D=10\,\rm Mpc$. These latter curves are cut off at the frequency of the Innermost Stable Circular Orbit $f\lesssim1\,{\rm kHz}\,({\rm M_\odot}/M_c)$ evaluated for equal-mass binaries \citep{MaggioreMichele2008GWV1}, above which the SMBHBs quickly merge. We also show sensitivity curves from \citet[][Figure 5]{2022PhRvD.105d4005B}, as well as the sensitivity curves of Pulsar Timing Arrays (PPTA \citep[][]{2010MNRAS.407..669Y}; EPTA \citep[][]{2016MNRAS.455.1665B}; NANOGrav \citep[][]{2019ApJ...880..116A}). The detection threshold ($\rm SNR=2$) is chosen to allow a consistent comparison to reported PTA sensitivities. In practice, we expect our method to show a reduction in sensitivity around $F\simeq1/{\rm yr}\simeq32\,\rm nHz$ as seen for PTAs, where it would be difficult to distinguish a background GW from the Doppler modulation due the annual motion of LISA around the sun. The sensitivity of our method is limited to frequencies $F\gtrsim 1/T$ (e.g., $32\,\rm nHz$ in the {\tt Pessimistic} model), below which the sensitivity is limited by the finite width of the frequency bins.}
 \label{fig:A-F}
\end{figure*}
We estimate the sensitivity to background gravitational waves for the three models using our method, as in Eq.~\eqref{eq:snr}. Figure~\ref{fig:A-F} shows the amplitude $A$ versus frequency $F$ of a background GW that could be detected with $\rm SNR=2$, corresponding to a $\simeq95\,\%$ detection probability. The differences between the $\tt Pessimistic$ and $\tt Optimistic$ models are less than one order of magnitude in $A$. 
Our method is sensitive to GWs with frequencies as low as $F\sim10^{-8}\,\rm Hz$. GWs of these frequencies could be present in our Universe, e.g., as part of a (stochastic) background of GWs emitted by numerous individual sources \citep{2008MNRAS.390..192S}. At a frequency of $F\simeq10^{-8}\,\rm Hz$ our method would be sensitive to amplitudes $A\gtrsim10^{-10}$; GWs of that amplitude at that frequency could, e.g., be emitted by a very massive SMBHB with a chirp mass of several $\sim10^{10}\,\rm M_\odot$ at a distance $D=10\,\rm Mpc$, which is the scale of the Virgo cluster. The existence of such an object is unlikely, since observations from pulsar timing indicate that there is no SMBHB of that mass within a distance $D\lesssim\mathcal{O}(1)\,\rm Gpc$ \citep{2019ApJ...880..116A}. Yet, analysing the GWs from BWDs could pose an independent method to probe the existence of these very massive SMBHBs.  

\begin{figure}
 \includegraphics[width=1\columnwidth]{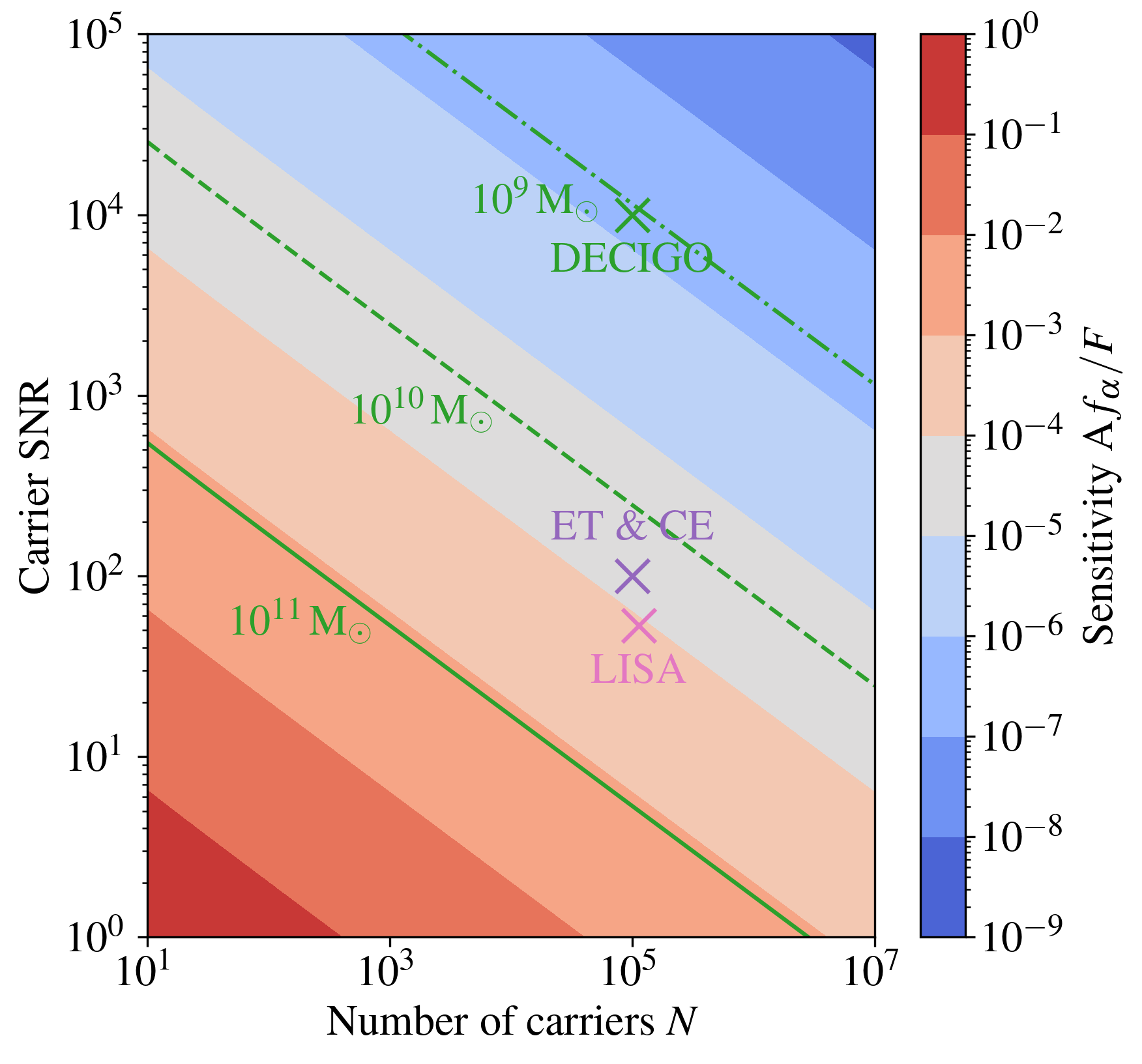}
 \caption{Order-of-magnitude estimate for the sensitivity to background gravitational waves (GWs) by cross-correlating a generic set of a number of GW signals $N$ that are each detected with a certain SNR (`Carrier SNR'). The sensitivity (given by the colour scale) is expressed as the product of the background amplitude $A$ times the typical frequency ratio of the background and carrier signals $f_\alpha/F$, where the detection threshold corresponds to an $\rm SNR$ equal to one.
Furthermore, we indicate the sensitivity that could be obtained using a set of GW signals in the $\rm dHz$ regime from binary neutron stars as carriers, which could be done using data from DECIGO \citep[][]{2001PhRvL..87v1103S}, and similarly the sensitivity using carrier signals detected using ET and CE \citep{2020JCAP...03..050M}. We also show the sensitivity that could be obtained using the average SNR of binary white dwarf signals detected by LISA (in the {\tt Moderate} model), as explicated in Figure~\ref{fig:A-F}. For these detectors we assume typical carrier frequencies of $f_\alpha\simeq0.1\,\rm Hz$ (DECIGO), $10\,\rm Hz$ (ET/CE), and $10^{-3}\,\rm Hz$ (LISA). For reference, we show contour lines that correspond to GW amplitudes from super-massive binary black holes with chirp masses ranging from $10^{9}$ to $10^{11}\,\rm M_\odot$ at a fiducial distance $D=10\,\rm Mpc$, with a background frequency $F=10^{-8}\,\rm Hz$, and a carrier frequency $f_\alpha=0.1\,\rm Hz$.}
 \label{fig:n-snr}
\end{figure}

We also consider the more general case of a number of carrier GW signals observed with an unspecified GW detector. For this case we assume that all $N$ carrier signals have a similar frequency and are detected with the same SNR~$\sim a_\alpha \sqrt{T/\rho(f_\alpha)}={\rm const.}$. In Figure~\ref{fig:n-snr}, we show the correlated background GW amplitude that can be detected at an SNR of one, as a function of the number and individual SNR of the carrier signals. 

We can apply this result to a proposed next-generation GW detector such as DECIGO \citep[][]{2001PhRvL..87v1103S,2011PhRvD..83d4011Y,2018PTEP.2018g3E01I}, which operates in the $\rm dHz$ regime and is expected to observe GWs from a large number of compact binary stars. 
Assuming DECIGO observes GW signals from a population of $N=10^5$ binary neutron stars (BNSs) each observed with an SNR of $\sim10^4$ \citep[][]{2001PhRvL..87v1103S} at a typical frequency of $f_\alpha=0.1\,\rm Hz$, it would be possible to detect background GWs from SMBHBs with chirp masses of about $\sim10^9\,\rm M_\odot$ (at a fiducial distance $D=10\,\rm Mpc$ and frequency $F=10^{-8}\,\rm Hz$). This would make the sensitivity of DECIGO to low-frequency GWs competitive with that of current PTAs (cf. Figure~\ref{fig:A-F} and \citep[][]{2023arXiv231106335S}).

For reference, we also indicate in Figure~\ref{fig:n-snr} the sensitivity that could be obtained using $\sim10^5$ carrier signals with an SNR~$\sim10^2$ from compact binary coalescences, as expected to be detected using both Einstein Telescope (ET) and Cosmic Explorer (CE) \citep{2020JCAP...03..050M}. These carrier signals would have frequencies between $10$ and $10^3$~Hz and could be observed for a duration $T\lesssim10^3$~s, which means the minimum detectable background GW frequency using our method is $F\sim10^{-3}\,\rm Hz$. Coherent background GW signals may be searched for using non-coincident carrier signals with a slight modification of the method described in Sec.~\ref{sec:data-analysis}; a frequency-dependent phase correction ($\phi_\mathrm{corr}=2\pi T_\mathrm{diff}F$) must be applied to each carrier's demodulated spectrum (Eq.~\ref{eq:carrierspectrum}), for a time difference between the signals $T_\mathrm{diff}$. In case the background GW signal has a coherence time much shorter than the total observation time for all signals (i.e., the detector's lifetime), only coincident carrier signals can be cross-correlated to gain sensitivity.

The sensitivity of our method is fundamentally limited to frequencies $F\gtrsim 1/T$, as for lower frequencies the background signal cannot be distinguished from the carrier \citep{2022PhRvD.105d4005B}. The same low-frequency limit due to observation time exists for PTAs. The high-frequency limit of our method is set by the Nyquist frequency of the detector output sampling, $f_s/2$, where for LISA $f_s\sim1\,\rm Hz\;$ \citep[][]{2022PhRvD.105d4005B}. PTAs have a much smaller sensitive bandwidth due to the low observation cadence of radio telescopes (once every several days or less).

\section{Discussion}\label{sec:discussion}
An important assumption used in deriving the sensitivity projections above is that the carrier signal frequency, amplitude, and phase are known without uncertainty. In practice, there will be non-negligible uncertainty in the determination of these parameters for the BWD signals detected with LISA, which means the normalisation and demodulation in Eq.~\ref{eq:demodulation} is subject to inaccuracy. The error in performing this operation on the data would propagate to an inaccuracy in the addition of carrier sidebands across different spectra (i.e. different sidebands would not line up in frequency exactly and their coherent summation would be imperfect). This uncertainty in the carrier GW parameters would thus reduce the signal-to-noise ratio of background GW signals, and needs to be accounted for in Eq.~\ref{eq:snr}, which we leave for future work.

An effect of particular concern that manifests as uncertainty of the frequency of the carrier signal is phase noise imparted by the data acquisition system of the gravitational-wave detector. As this noise would appear as random modulations of the carrier signal, it could obfuscate any background GWs that produce the same effect. Phase noise in the data acquisition system, due to, e.g., timing jitter of the sampling clocks, would produce irreducible correlated noise in the demodulated cross-spectra of different carriers. This effect might only be reduced by cross-correlating data obtained with different uncorrelated oscillators. Similarly, stochastic phase noise intrinsic to the carrier GW signal would reduce sensitivity to background GWs. In this case the effect on the sensitivity is limited as this noise will be uncorrelated between carriers and will be reduced in the average cross-spectrum (Eq.~\ref{eq:avg_cross_spectrum}). 

\begin{figure}
 \includegraphics[width=1\columnwidth]{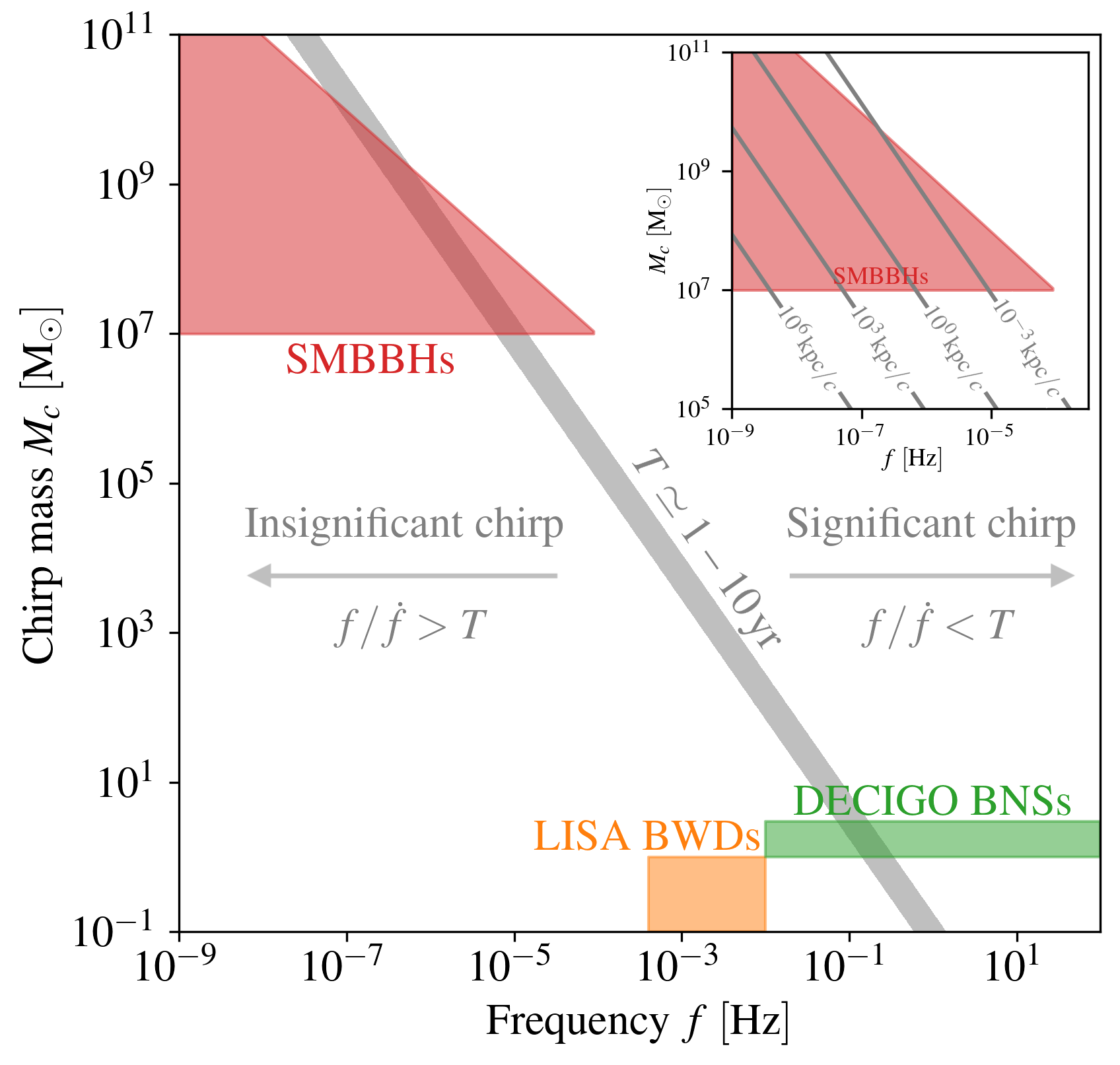}
 \caption{Timescale $f/\dot{f}=(5/96)(c^3/GM_c)^{5/3}(\pi f)^{-8/3}$ at which the frequency $f$ of a compact binary with chirp mass $M_c$ significantly increases due to energy loss through gravitational-wave emission. Coloured boxes indicate the parameter regions of background super-massive binary black holes (SMBHBs), LISA binary white dwarfs (BWDs), and DECIGO binary neutron stars (BNSs). This shows that LISA BWDs and most of the SMBHBs would not undergo significant frequency changes within the observation time $T\simeq1$~--~$10\,\rm yr$, whereas most DECIGO BNSs would. The inset shows whether the SMBHBs would exhibit significant frequency changes within typical light travel times between a carrier source and the observer, i.e., whether `local' and `distant' sidebands overlap or not. For this figure we take the maximum GW frequency emitted by SMBHBs to correspond to the Innermost Stable Circular Orbit $f\lesssim1\,{\rm kHz}\,({\rm M_\odot}/M_c)$ evaluated for equal-mass binaries \citep{MaggioreMichele2008GWV1}, which causes the diagonal cut-off. }
 \label{fig:chirping-timescale}
\end{figure}

In addition to these effective stochastic fluctuations of the carrier signal, there could be deterministic frequency changes of the carrier and background GWs. If the frequency of the background GWs changes significantly over the measurement time, i.e., if the GW background power spectral density is non-stationary, the coherent signal power would be spread over multiple frequency bins, leading to a lower SNR in each bin. An SMBHB background source might undergo a significant frequency evolution as its orbital period decays due to energy loss by GW emission. Figure~\ref{fig:chirping-timescale} shows that this frequency change $\dot{F}$ (`chirp') would not be significant for SMBHBs ($M_c\gtrsim10^9$) over the duration of observation $T\simeq1$~--~$10\,\rm yr$. Figure~\ref{fig:chirping-timescale} also shows the expected frequency changes of the LISA and DECIGO carrier signals. In particular, it shows that most DECIGO BNSs undergo significant frequency evolution over the duration of the detected signal. As discussed in Sec.~\ref{sec:data-analysis}, these frequency changes could be compensated for at the demodulation stage. 

Non-stationarity of the background GW PSD has another effect; the frequency change over a time equal to the typical light travel time between the carrier source and observer determines the frequency-space separation of the `local' and `distant' sideband terms, i.e., ${|F_L-F_D| \propto d_\alpha\dot{F}/c}$. If these terms are not separated in the spectrum, i.e., when $|F_L-F_D|\lesssim1/T$, coherent summation of the `local' terms of different cross-spectra is still possible but the `distant' terms would add a small incoherent noise-like contribution to any signal bin. The inset of Figure~\ref{fig:chirping-timescale} shows that given typical light travel times between BWDs and the LISA detector of $d_\alpha/c\simeq10^{-1}$~--~$10^{1}\,{\rm kpc}/c$ \citep{2019MNRAS.490.5888L}, both separated and non-separated sidebands could be observed for background SMBHB GW sources. On the other hand, DECIGO will observe carrier signals from BNSs at much larger distances, e.g., $d_\alpha\simeq10^4\,{\rm kpc}$ for a GW170817-like event \citep[][]{2017ApJ...848L..12A}, and therefore `local' and 'distant' sidebands produced by a background SMBHB source ($M_c\gtrsim10^9\,\rm M_\odot$) would be well-separated in DECIGO data.

\begin{figure}
 \includegraphics[width=1\columnwidth]{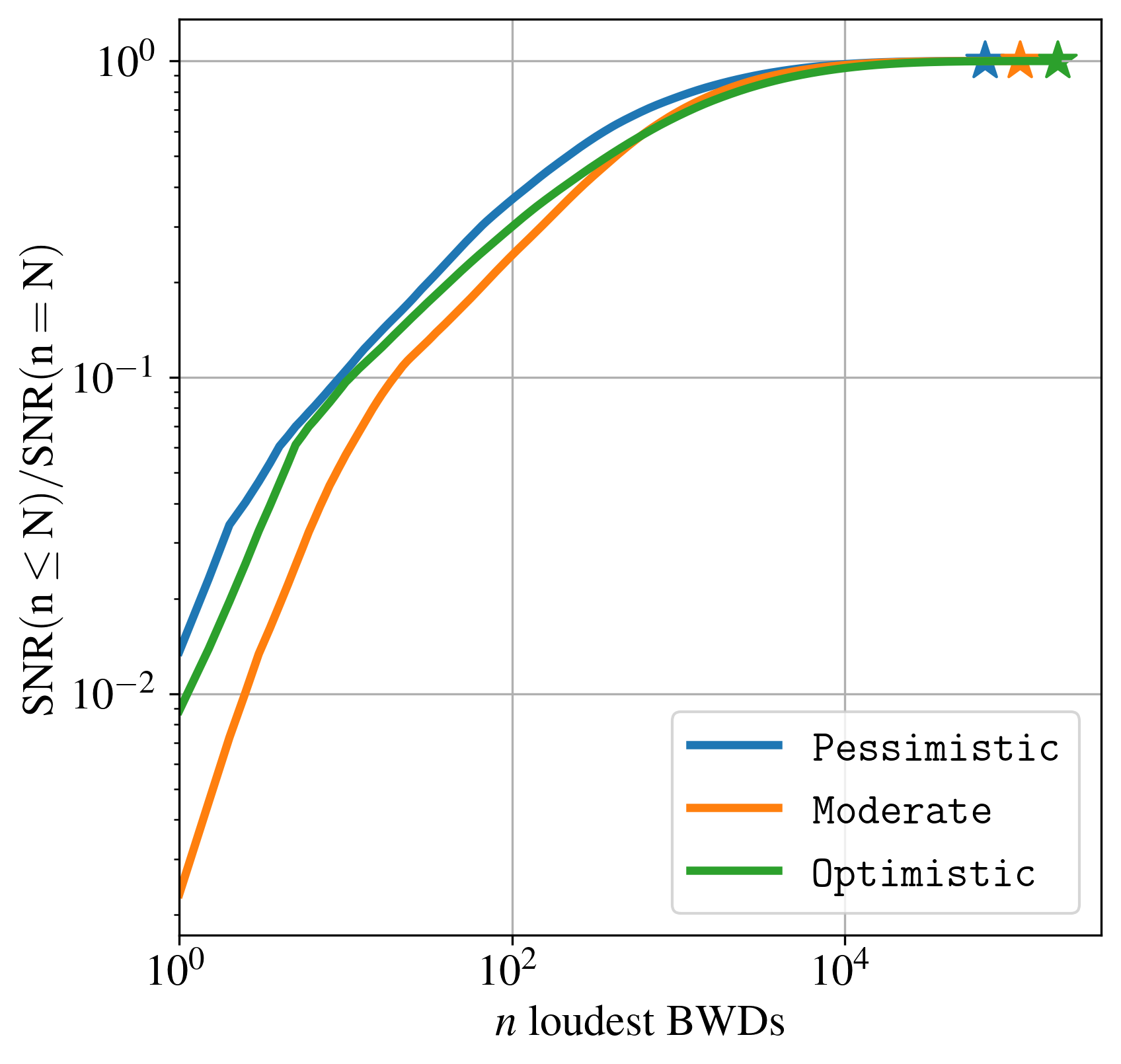}
 \caption{Cumulative $\rm SNR$ of a background gravitational wave signal as a function of the $n$ loudest binary white dwarfs (BWDs) in the set of carrier signals. 
 Stars at the end of each line indicate the total number $N$ of binaries in each model. In any model several $10^2$ to $10^3$ of the loudest BWDs are enough to achieve sensitivities similar to the entire sample.}
 \label{fig:cumulative}
\end{figure}
We note that for the sensitivity projections for LISA, the number $N$ of individually resolvable BWDs in our models (see Table~\ref{tab:parameters}) is larger by a factor up to $\sim 10$ compared to previous estimates from Galaxy models combined with a binary population model \citep[][]{2001A&A...368..939N,2001A&A...375..890N,2010ApJ...717.1006R,2012ApJ...758..131N,2019MNRAS.490.5888L}, which reflects the large uncertainty of current predictions about the detectable BWD population. However, the exact total number of BWDs does not significantly affect the estimated sensitivity because the $\sim\mathcal{O}(10^3)$ loudest BWDs signals provide the dominant contribution to the sensitivity. This is shown in Figure~\ref{fig:cumulative}; where we plot the normalised cumulative contribution of BWDs to the total $\rm SNR$.
It can be seen that several $10^2$ to $10^3$ BWDs are enough to achieve similar sensitivities to the total BWD population.

Importantly, we note that the estimates for detecting low-frequency gravitational waves using BWD carrier signals in LISA as obtained in \citet{2022PhRvD.105d4005B} yield a sensitivity that is worse than the sensitivity we project for this scenario by roughly an order of magnitude (for the {\tt moderate} model, see Figure~\ref{fig:A-F}). \citet{2022PhRvD.105d4005B} obtain their main sensitivity estimate for the background GW amplitude by evaluating the Fisher information, which is equivalent to the maximum likelihood estimator in our Eq.~\ref{eq:snr}. However, \citet{2022PhRvD.105d4005B} make different assumptions about the carrier signals and their uncertainty. The lower sensitivity estimate found in their work is likely in part due to the inclusion of the uncertainty in the estimation of the carrier signal parameters, and the Fisher-information approach therefore presents a theoretical upper bound on the sensitivity that can be obtained with our method. Our estimate does not account for uncertainty in estimating the carrier signal parameters as explained above, whereby we potentially overestimate the sensitivity of our method. 
Another aspect that could account for some of the discrepancy between our sensitivity projections and those of \citet{2022PhRvD.105d4005B} is the difference in BWD carrier signal populations used; the numbers of individually resolvable BWDs in the observationally-driven populations used in our work is greater than the number assumed in \citet{2022PhRvD.105d4005B}.

\section{Conclusion}
In this work, we have outlined a method to use a set of carrier gravitational wave sources to search for correlated frequency modulations caused by low-frequency background GWs. In this method demodulated cross-spectra of carrier sources are added coherently and with optimal weights such that any modulation common to the carrier sources is amplified with respect to random detector noise. 

We considered the case of using our method to search for low-frequency GWs in data from LISA, which is expected to detect GWs from a large number of Galactic binary white dwarfs. The projected sensitivity that could thus be obtained (Figure~\ref{fig:A-F}) ranges from strain amplitudes of $A\sim10^{-10}$ at $F\sim10^{-8}$~Hz to $\sim10^{-7}$ at $\sim10^{-5}$~Hz, and would cover a part of the GW spectrum where no other detection methods are currently available. 

This sensitivity could potentially enable the detection of very massive SMBHBs with a chirp mass of several $10^{10}\,\rm M_\odot$ at a distance of $D=10\,\rm Mpc$, although the existence of such a system is all but ruled out by pulsar timing array observations~\citep{2019ApJ...880..116A}.

Our results show that an even better sensitivity could be achieved using GW signals from compact binary stars detectable with next-generation GW detectors that operate in the $\rm dHz$ regime. In particular, using signals of binary neutron stars expected to be detected with DECIGO would yield a sensitivity competitive with that of current pulsar timing arrays.

Future detectors designed to detect GW signals in a higher frequency range could be used to indirectly probe GWs down to the frequencies given by the inverse instrument lifetime. Potentially, this allows one to probe GW frequencies in the $\rm nHz$~--~$\mu\rm Hz$ range, which are expected to be emitted by the most massive black holes in our Universe. Conveniently, this could be achieved without modification of the detector designs and with the same data outputs. This method could therefore prove a valuable tool in the exploration of the gravitational-wave spectrum  and the development of gravitational-wave astronomy in general.

\section*{Acknowledgements}
We thank Vivien Raymond, Bangalore Sathyaprakash, Fabio Antonini, Hartmut Grote, Guido M\"uller, Antoine Petiteau, Martin Hewitson, Lucio Mayer, and Valeriya Korol for helpful input and discussions. 

J.S.~acknowledges funding from the Netherlands Organisation for Scientific Research (NWO), as part of the Vidi research program BinWaves (project number 639.042.728, PI: de Mink). S.M.V.~acknowledges funding from the Leverhulme Trust in the UK under research grant RPG-2019-022. 

\section*{Data Availability}
The data underlying this article will be shared on reasonable request to the authors.

\bibliography{ms}

\begin{thebibliography}{57}%
\makeatletter
\providecommand \@ifxundefined [1]{%
 \@ifx{#1\undefined}
}%
\providecommand \@ifnum [1]{%
 \ifnum #1\expandafter \@firstoftwo
 \else \expandafter \@secondoftwo
 \fi
}%
\providecommand \@ifx [1]{%
 \ifx #1\expandafter \@firstoftwo
 \else \expandafter \@secondoftwo
 \fi
}%
\providecommand \natexlab [1]{#1}%
\providecommand \enquote  [1]{``#1''}%
\providecommand \bibnamefont  [1]{#1}%
\providecommand \bibfnamefont [1]{#1}%
\providecommand \citenamefont [1]{#1}%
\providecommand \href@noop [0]{\@secondoftwo}%
\providecommand \href [0]{\begingroup \@sanitize@url \@href}%
\providecommand \@href[1]{\@@startlink{#1}\@@href}%
\providecommand \@@href[1]{\endgroup#1\@@endlink}%
\providecommand \@sanitize@url [0]{\catcode `\\12\catcode `\$12\catcode
  `\&12\catcode `\#12\catcode `\^12\catcode `\_12\catcode `\%12\relax}%
\providecommand \@@startlink[1]{}%
\providecommand \@@endlink[0]{}%
\providecommand \url  [0]{\begingroup\@sanitize@url \@url }%
\providecommand \@url [1]{\endgroup\@href {#1}{\urlprefix }}%
\providecommand \urlprefix  [0]{URL }%
\providecommand \Eprint [0]{\href }%
\providecommand \doibase [0]{https://doi.org/}%
\providecommand \selectlanguage [0]{\@gobble}%
\providecommand \bibinfo  [0]{\@secondoftwo}%
\providecommand \bibfield  [0]{\@secondoftwo}%
\providecommand \translation [1]{[#1]}%
\providecommand \BibitemOpen [0]{}%
\providecommand \bibitemStop [0]{}%
\providecommand \bibitemNoStop [0]{.\EOS\space}%
\providecommand \EOS [0]{\spacefactor3000\relax}%
\providecommand \BibitemShut  [1]{\csname bibitem#1\endcsname}%
\let\auto@bib@innerbib\@empty
\bibitem [{\citenamefont {{Abbott, B.~P. et al.}}(2016)}]{GW150914}%
  \BibitemOpen
  \bibfield  {author} {\bibinfo {author} {\bibnamefont {{Abbott, B.~P. et
  al.}}},\ }\bibfield  {title} {\bibinfo {title} {{Observation of Gravitational
  Waves from a Binary Black Hole Merger}},\ }\href
  {https://doi.org/10.1103/PhysRevLett.116.061102} {\bibfield  {journal}
  {\bibinfo  {journal} {\prl}\ }\textbf {\bibinfo {volume} {116}},\ \bibinfo
  {eid} {061102} (\bibinfo {year} {2016})},\ \Eprint
  {https://arxiv.org/abs/1602.03837} {arXiv:1602.03837 [gr-qc]} \BibitemShut
  {NoStop}%
\bibitem [{\citenamefont {{The LIGO Scientific
  Collaboration}}(2021)}]{2021arXiv211103634T}%
  \BibitemOpen
  \bibfield  {author} {\bibinfo {author} {\bibnamefont {{The LIGO Scientific
  Collaboration}}},\ }\bibfield  {title} {\bibinfo {title} {{The population of
  merging compact binaries inferred using gravitational waves through
  GWTC-3}},\ }\href@noop {} {\bibfield  {journal} {\bibinfo  {journal} {arXiv
  e-prints}\ ,\ \bibinfo {eid} {arXiv:2111.03634}} (\bibinfo {year} {2021})},\
  \Eprint {https://arxiv.org/abs/2111.03634} {arXiv:2111.03634 [astro-ph.HE]}
  \BibitemShut {NoStop}%
\bibitem [{\citenamefont {{Owen}}\ \emph {et~al.}(1998)\citenamefont {{Owen}},
  \citenamefont {{Lindblom}}, \citenamefont {{Cutler}}, \citenamefont
  {{Schutz}}, \citenamefont {{Vecchio}},\ and\ \citenamefont
  {{Andersson}}}]{1998PhRvD..58h4020O}%
  \BibitemOpen
  \bibfield  {author} {\bibinfo {author} {\bibfnamefont {B.~J.}\ \bibnamefont
  {{Owen}}}, \bibinfo {author} {\bibfnamefont {L.}~\bibnamefont {{Lindblom}}},
  \bibinfo {author} {\bibfnamefont {C.}~\bibnamefont {{Cutler}}}, \bibinfo
  {author} {\bibfnamefont {B.~F.}\ \bibnamefont {{Schutz}}}, \bibinfo {author}
  {\bibfnamefont {A.}~\bibnamefont {{Vecchio}}},\ and\ \bibinfo {author}
  {\bibfnamefont {N.}~\bibnamefont {{Andersson}}},\ }\bibfield  {title}
  {\bibinfo {title} {{Gravitational waves from hot young rapidly rotating
  neutron stars}},\ }\href {https://doi.org/10.1103/PhysRevD.58.084020}
  {\bibfield  {journal} {\bibinfo  {journal} {Physical Review D}\ }\textbf
  {\bibinfo {volume} {58}},\ \bibinfo {eid} {084020} (\bibinfo {year}
  {1998})},\ \Eprint {https://arxiv.org/abs/gr-qc/9804044} {arXiv:gr-qc/9804044
  [gr-qc]} \BibitemShut {NoStop}%
\bibitem [{\citenamefont {{Nelemans}}\ \emph
  {et~al.}(2001{\natexlab{a}})\citenamefont {{Nelemans}}, \citenamefont
  {{Yungelson}},\ and\ \citenamefont {{Portegies
  Zwart}}}]{2001A&A...375..890N}%
  \BibitemOpen
  \bibfield  {author} {\bibinfo {author} {\bibfnamefont {G.}~\bibnamefont
  {{Nelemans}}}, \bibinfo {author} {\bibfnamefont {L.~R.}\ \bibnamefont
  {{Yungelson}}},\ and\ \bibinfo {author} {\bibfnamefont {S.~F.}\ \bibnamefont
  {{Portegies Zwart}}},\ }\bibfield  {title} {\bibinfo {title} {{The
  gravitational wave signal from the Galactic disk population of binaries
  containing two compact objects}},\ }\href
  {https://doi.org/10.1051/0004-6361:20010683} {\bibfield  {journal} {\bibinfo
  {journal} {Astronomy \& Astrophysics}\ }\textbf {\bibinfo {volume} {375}},\
  \bibinfo {pages} {890} (\bibinfo {year} {2001}{\natexlab{a}})},\ \Eprint
  {https://arxiv.org/abs/astro-ph/0105221} {arXiv:astro-ph/0105221 [astro-ph]}
  \BibitemShut {NoStop}%
\bibitem [{\citenamefont {{Klein}}\ \emph {et~al.}(2016)\citenamefont
  {{Klein}}, \citenamefont {{Barausse}}, \citenamefont {{Sesana}},
  \citenamefont {{Petiteau}}, \citenamefont {{Berti}}, \citenamefont {{Babak}},
  \citenamefont {{Gair}}, \citenamefont {{Aoudia}}, \citenamefont {{Hinder}},
  \citenamefont {{Ohme}},\ and\ \citenamefont
  {{Wardell}}}]{2016PhRvD..93b4003K}%
  \BibitemOpen
  \bibfield  {author} {\bibinfo {author} {\bibfnamefont {A.}~\bibnamefont
  {{Klein}}}, \bibinfo {author} {\bibfnamefont {E.}~\bibnamefont {{Barausse}}},
  \bibinfo {author} {\bibfnamefont {A.}~\bibnamefont {{Sesana}}}, \bibinfo
  {author} {\bibfnamefont {A.}~\bibnamefont {{Petiteau}}}, \bibinfo {author}
  {\bibfnamefont {E.}~\bibnamefont {{Berti}}}, \bibinfo {author} {\bibfnamefont
  {S.}~\bibnamefont {{Babak}}}, \bibinfo {author} {\bibfnamefont
  {J.}~\bibnamefont {{Gair}}}, \bibinfo {author} {\bibfnamefont
  {S.}~\bibnamefont {{Aoudia}}}, \bibinfo {author} {\bibfnamefont
  {I.}~\bibnamefont {{Hinder}}}, \bibinfo {author} {\bibfnamefont
  {F.}~\bibnamefont {{Ohme}}},\ and\ \bibinfo {author} {\bibfnamefont
  {B.}~\bibnamefont {{Wardell}}},\ }\bibfield  {title} {\bibinfo {title}
  {{Science with the space-based interferometer eLISA: Supermassive black hole
  binaries}},\ }\href {https://doi.org/10.1103/PhysRevD.93.024003} {\bibfield
  {journal} {\bibinfo  {journal} {Physical Review D}\ }\textbf {\bibinfo
  {volume} {93}},\ \bibinfo {eid} {024003} (\bibinfo {year} {2016})},\ \Eprint
  {https://arxiv.org/abs/1511.05581} {arXiv:1511.05581 [gr-qc]} \BibitemShut
  {NoStop}%
\bibitem [{\citenamefont {{Mandic}}\ and\ \citenamefont
  {{Buonanno}}(2006)}]{2006PhRvD..73f3008M}%
  \BibitemOpen
  \bibfield  {author} {\bibinfo {author} {\bibfnamefont {V.}~\bibnamefont
  {{Mandic}}}\ and\ \bibinfo {author} {\bibfnamefont {A.}~\bibnamefont
  {{Buonanno}}},\ }\bibfield  {title} {\bibinfo {title} {{Accessibility of the
  pre-big-bang models to LIGO}},\ }\href
  {https://doi.org/10.1103/PhysRevD.73.063008} {\bibfield  {journal} {\bibinfo
  {journal} {Physical Review D}\ }\textbf {\bibinfo {volume} {73}},\ \bibinfo
  {eid} {063008} (\bibinfo {year} {2006})},\ \Eprint
  {https://arxiv.org/abs/astro-ph/0510341} {arXiv:astro-ph/0510341 [astro-ph]}
  \BibitemShut {NoStop}%
\bibitem [{\citenamefont {{Carr}}\ \emph {et~al.}(2016)\citenamefont {{Carr}},
  \citenamefont {{K{\"u}hnel}},\ and\ \citenamefont
  {{Sandstad}}}]{2016PhRvD..94h3504C}%
  \BibitemOpen
  \bibfield  {author} {\bibinfo {author} {\bibfnamefont {B.}~\bibnamefont
  {{Carr}}}, \bibinfo {author} {\bibfnamefont {F.}~\bibnamefont
  {{K{\"u}hnel}}},\ and\ \bibinfo {author} {\bibfnamefont {M.}~\bibnamefont
  {{Sandstad}}},\ }\bibfield  {title} {\bibinfo {title} {{Primordial black
  holes as dark matter}},\ }\href {https://doi.org/10.1103/PhysRevD.94.083504}
  {\bibfield  {journal} {\bibinfo  {journal} {Physical Review D}\ }\textbf
  {\bibinfo {volume} {94}},\ \bibinfo {eid} {083504} (\bibinfo {year}
  {2016})},\ \Eprint {https://arxiv.org/abs/1607.06077} {arXiv:1607.06077
  [astro-ph.CO]} \BibitemShut {NoStop}%
\bibitem [{\citenamefont {Brito}\ \emph {et~al.}(2017)\citenamefont {Brito},
  \citenamefont {Ghosh}, \citenamefont {Barausse}, \citenamefont {Berti},
  \citenamefont {Cardoso}, \citenamefont {Dvorkin}, \citenamefont {Klein},\
  and\ \citenamefont {Pani}}]{brito_gravitational_2017}%
  \BibitemOpen
  \bibfield  {author} {\bibinfo {author} {\bibfnamefont {R.}~\bibnamefont
  {Brito}}, \bibinfo {author} {\bibfnamefont {S.}~\bibnamefont {Ghosh}},
  \bibinfo {author} {\bibfnamefont {E.}~\bibnamefont {Barausse}}, \bibinfo
  {author} {\bibfnamefont {E.}~\bibnamefont {Berti}}, \bibinfo {author}
  {\bibfnamefont {V.}~\bibnamefont {Cardoso}}, \bibinfo {author} {\bibfnamefont
  {I.}~\bibnamefont {Dvorkin}}, \bibinfo {author} {\bibfnamefont
  {A.}~\bibnamefont {Klein}},\ and\ \bibinfo {author} {\bibfnamefont
  {P.}~\bibnamefont {Pani}},\ }\bibfield  {title} {\bibinfo {title}
  {Gravitational wave searches for ultralight bosons with {LIGO} and {LISA}},\
  }\bibfield  {journal} {\bibinfo  {journal} {Physical Review D}\ }\textbf
  {\bibinfo {volume} {96}},\ \href {https://doi.org/10.1103/PhysRevD.96.064050}
  {10.1103/PhysRevD.96.064050} (\bibinfo {year} {2017}),\ \bibinfo {note}
  {arXiv: 1706.06311}\BibitemShut {NoStop}%
\bibitem [{\citenamefont {Vermeulen}\ \emph {et~al.}(2021)\citenamefont
  {Vermeulen}, \citenamefont {Aiello}, \citenamefont {Ejlli}, \citenamefont
  {Griffiths}, \citenamefont {James}, \citenamefont {Dooley},\ and\
  \citenamefont {Grote}}]{vermeulen_experiment_2021}%
  \BibitemOpen
  \bibfield  {author} {\bibinfo {author} {\bibfnamefont {S.~M.}\ \bibnamefont
  {Vermeulen}}, \bibinfo {author} {\bibfnamefont {L.}~\bibnamefont {Aiello}},
  \bibinfo {author} {\bibfnamefont {A.}~\bibnamefont {Ejlli}}, \bibinfo
  {author} {\bibfnamefont {W.~L.}\ \bibnamefont {Griffiths}}, \bibinfo {author}
  {\bibfnamefont {A.~L.}\ \bibnamefont {James}}, \bibinfo {author}
  {\bibfnamefont {K.~L.}\ \bibnamefont {Dooley}},\ and\ \bibinfo {author}
  {\bibfnamefont {H.}~\bibnamefont {Grote}},\ }\bibfield  {title} {\bibinfo
  {title} {An experiment for observing quantum gravity phenomena using twin
  table-top {3D} interferometers},\ }\href
  {https://doi.org/10.1088/1361-6382/abe757} {\bibfield  {journal} {\bibinfo
  {journal} {Classical and Quantum Gravity}\ }\textbf {\bibinfo {volume}
  {38}},\ \bibinfo {pages} {085008} (\bibinfo {year} {2021})}\BibitemShut
  {NoStop}%
\bibitem [{\citenamefont {Chou}\ \emph {et~al.}(2017)\citenamefont {Chou},
  \citenamefont {Glass}, \citenamefont {Gustafson}, \citenamefont {Hogan},
  \citenamefont {Kamai}, \citenamefont {Kwon}, \citenamefont {Lanza},
  \citenamefont {McCuller}, \citenamefont {Meyer}, \citenamefont {Richardson},
  \citenamefont {Stoughton}, \citenamefont {Tomlin},\ and\ \citenamefont
  {Weiss}}]{chou_holometer:_2017}%
  \BibitemOpen
  \bibfield  {author} {\bibinfo {author} {\bibfnamefont {A.}~\bibnamefont
  {Chou}}, \bibinfo {author} {\bibfnamefont {H.}~\bibnamefont {Glass}},
  \bibinfo {author} {\bibfnamefont {H.~R.}\ \bibnamefont {Gustafson}}, \bibinfo
  {author} {\bibfnamefont {C.}~\bibnamefont {Hogan}}, \bibinfo {author}
  {\bibfnamefont {B.~L.}\ \bibnamefont {Kamai}}, \bibinfo {author}
  {\bibfnamefont {O.}~\bibnamefont {Kwon}}, \bibinfo {author} {\bibfnamefont
  {R.}~\bibnamefont {Lanza}}, \bibinfo {author} {\bibfnamefont
  {L.}~\bibnamefont {McCuller}}, \bibinfo {author} {\bibfnamefont {S.~S.}\
  \bibnamefont {Meyer}}, \bibinfo {author} {\bibfnamefont {J.}~\bibnamefont
  {Richardson}}, \bibinfo {author} {\bibfnamefont {C.}~\bibnamefont
  {Stoughton}}, \bibinfo {author} {\bibfnamefont {R.}~\bibnamefont {Tomlin}},\
  and\ \bibinfo {author} {\bibfnamefont {R.}~\bibnamefont {Weiss}},\ }\bibfield
   {title} {\bibinfo {title} {The {Holometer}: {An} {Instrument} to {Probe}
  {Planckian} {Quantum} {Geometry}},\ }\href
  {https://doi.org/10.1088/1361-6382/aa5e5c} {\bibfield  {journal} {\bibinfo
  {journal} {Classical and Quantum Gravity}\ }\textbf {\bibinfo {volume}
  {34}},\ \bibinfo {pages} {065005} (\bibinfo {year} {2017})},\ \bibinfo {note}
  {arXiv: 1611.08265}\BibitemShut {NoStop}%
\bibitem [{\citenamefont {{P. Amaro-Seoane et
  al.}}(2017)}]{2017arXiv170200786A}%
  \BibitemOpen
  \bibfield  {author} {\bibinfo {author} {\bibnamefont {{P. Amaro-Seoane et
  al.}}},\ }\bibfield  {title} {\bibinfo {title} {{Laser Interferometer Space
  Antenna}},\ }\href@noop {} {\bibfield  {journal} {\bibinfo  {journal} {arXiv
  e-prints}\ ,\ \bibinfo {eid} {arXiv:1702.00786}} (\bibinfo {year} {2017})},\
  \Eprint {https://arxiv.org/abs/1702.00786} {arXiv:1702.00786 [astro-ph.IM]}
  \BibitemShut {NoStop}%
\bibitem [{\citenamefont {{A. Buikema et
  al.}}(2020)}]{buikema_sensitivity_2020}%
  \BibitemOpen
  \bibfield  {author} {\bibinfo {author} {\bibnamefont {{A. Buikema et al.}}},\
  }\bibfield  {title} {\bibinfo {title} {Sensitivity and performance of the
  advanced ligo detectors in the third observing run},\ }\href
  {https://doi.org/10.1103/PhysRevD.102.062003} {\bibfield  {journal} {\bibinfo
   {journal} {Physical Review D}\ }\textbf {\bibinfo {volume} {102}},\ \bibinfo
  {pages} {062003} (\bibinfo {year} {2020})}\BibitemShut {NoStop}%
\bibitem [{\citenamefont {Gertsenshtein}(1962)}]{gertsenshtein_wave_1962}%
  \BibitemOpen
  \bibfield  {author} {\bibinfo {author} {\bibfnamefont {M.~E.}\ \bibnamefont
  {Gertsenshtein}},\ }\bibfield  {title} {\bibinfo {title} {Wave resonance of
  light and gravitional waves},\ }\href@noop {} {\bibfield  {journal} {\bibinfo
   {journal} {Sov Phys JETP}\ }\textbf {\bibinfo {volume} {14}},\ \bibinfo
  {pages} {84} (\bibinfo {year} {1962})}\BibitemShut {NoStop}%
\bibitem [{\citenamefont {Ejlli}\ \emph {et~al.}(2019)\citenamefont {Ejlli},
  \citenamefont {Ejlli}, \citenamefont {Cruise}, \citenamefont {Pisano},\ and\
  \citenamefont {Grote}}]{ejlli_upper_2019}%
  \BibitemOpen
  \bibfield  {author} {\bibinfo {author} {\bibfnamefont {A.}~\bibnamefont
  {Ejlli}}, \bibinfo {author} {\bibfnamefont {D.}~\bibnamefont {Ejlli}},
  \bibinfo {author} {\bibfnamefont {A.~M.}\ \bibnamefont {Cruise}}, \bibinfo
  {author} {\bibfnamefont {G.}~\bibnamefont {Pisano}},\ and\ \bibinfo {author}
  {\bibfnamefont {H.}~\bibnamefont {Grote}},\ }\bibfield  {title} {\bibinfo
  {title} {Upper limits on the amplitude of ultra-high-frequency gravitational
  waves from graviton to photon conversion},\ }\href
  {https://doi.org/10.1140/epjc/s10052-019-7542-5} {\bibfield  {journal}
  {\bibinfo  {journal} {Eur. Phys. J. C}\ }\textbf {\bibinfo {volume} {79}},\
  \bibinfo {pages} {1032} (\bibinfo {year} {2019})},\ \bibinfo {note}
  {\_eprint: 1908.00232}\BibitemShut {NoStop}%
\bibitem [{\citenamefont {Goryachev}\ and\ \citenamefont
  {Tobar}(2014)}]{goryachev_gravitational_2014}%
  \BibitemOpen
  \bibfield  {author} {\bibinfo {author} {\bibfnamefont {M.}~\bibnamefont
  {Goryachev}}\ and\ \bibinfo {author} {\bibfnamefont {M.~E.}\ \bibnamefont
  {Tobar}},\ }\bibfield  {title} {\bibinfo {title} {Gravitational wave
  detection with high frequency phonon trapping acoustic cavities},\ }\href
  {https://doi.org/10.1103/PhysRevD.90.102005} {\bibfield  {journal} {\bibinfo
  {journal} {Physical Review D}\ }\textbf {\bibinfo {volume} {90}},\ \bibinfo
  {pages} {102005} (\bibinfo {year} {2014})},\ \bibinfo {note} {publisher:
  American Physical Society}\BibitemShut {NoStop}%
\bibitem [{\citenamefont {Aggarwal}\ \emph {et~al.}(2021)\citenamefont
  {Aggarwal}, \citenamefont {Aguiar}, \citenamefont {Bauswein}, \citenamefont
  {Cella}, \citenamefont {Clesse}, \citenamefont {Cruise}, \citenamefont
  {Domcke}, \citenamefont {Figueroa}, \citenamefont {Geraci}, \citenamefont
  {Goryachev}, \citenamefont {Grote}, \citenamefont {Hindmarsh}, \citenamefont
  {Muia}, \citenamefont {Mukund}, \citenamefont {Ottaway}, \citenamefont
  {Peloso}, \citenamefont {Quevedo}, \citenamefont {Ricciardone}, \citenamefont
  {Steinlechner}, \citenamefont {Steinlechner}, \citenamefont {Sun},
  \citenamefont {Tobar}, \citenamefont {Torrenti}, \citenamefont {Ünal},\ and\
  \citenamefont {White}}]{aggarwal_challenges_2021}%
  \BibitemOpen
  \bibfield  {author} {\bibinfo {author} {\bibfnamefont {N.}~\bibnamefont
  {Aggarwal}}, \bibinfo {author} {\bibfnamefont {O.~D.}\ \bibnamefont
  {Aguiar}}, \bibinfo {author} {\bibfnamefont {A.}~\bibnamefont {Bauswein}},
  \bibinfo {author} {\bibfnamefont {G.}~\bibnamefont {Cella}}, \bibinfo
  {author} {\bibfnamefont {S.}~\bibnamefont {Clesse}}, \bibinfo {author}
  {\bibfnamefont {A.~M.}\ \bibnamefont {Cruise}}, \bibinfo {author}
  {\bibfnamefont {V.}~\bibnamefont {Domcke}}, \bibinfo {author} {\bibfnamefont
  {D.~G.}\ \bibnamefont {Figueroa}}, \bibinfo {author} {\bibfnamefont
  {A.}~\bibnamefont {Geraci}}, \bibinfo {author} {\bibfnamefont
  {M.}~\bibnamefont {Goryachev}}, \bibinfo {author} {\bibfnamefont
  {H.}~\bibnamefont {Grote}}, \bibinfo {author} {\bibfnamefont
  {M.}~\bibnamefont {Hindmarsh}}, \bibinfo {author} {\bibfnamefont
  {F.}~\bibnamefont {Muia}}, \bibinfo {author} {\bibfnamefont {N.}~\bibnamefont
  {Mukund}}, \bibinfo {author} {\bibfnamefont {D.}~\bibnamefont {Ottaway}},
  \bibinfo {author} {\bibfnamefont {M.}~\bibnamefont {Peloso}}, \bibinfo
  {author} {\bibfnamefont {F.}~\bibnamefont {Quevedo}}, \bibinfo {author}
  {\bibfnamefont {A.}~\bibnamefont {Ricciardone}}, \bibinfo {author}
  {\bibfnamefont {J.}~\bibnamefont {Steinlechner}}, \bibinfo {author}
  {\bibfnamefont {S.}~\bibnamefont {Steinlechner}}, \bibinfo {author}
  {\bibfnamefont {S.}~\bibnamefont {Sun}}, \bibinfo {author} {\bibfnamefont
  {M.~E.}\ \bibnamefont {Tobar}}, \bibinfo {author} {\bibfnamefont
  {F.}~\bibnamefont {Torrenti}}, \bibinfo {author} {\bibfnamefont
  {C.}~\bibnamefont {Ünal}},\ and\ \bibinfo {author} {\bibfnamefont
  {G.}~\bibnamefont {White}},\ }\bibfield  {title} {\bibinfo {title}
  {Challenges and opportunities of gravitational-wave searches at {MHz} to
  {GHz} frequencies},\ }\href {https://doi.org/10.1007/s41114-021-00032-5}
  {\bibfield  {journal} {\bibinfo  {journal} {Living Reviews in Relativity}\
  }\textbf {\bibinfo {volume} {24}},\ \bibinfo {pages} {4} (\bibinfo {year}
  {2021})}\BibitemShut {NoStop}%
\bibitem [{\citenamefont {{Detweiler}}(1979)}]{1979ApJ...234.1100D}%
  \BibitemOpen
  \bibfield  {author} {\bibinfo {author} {\bibfnamefont {S.}~\bibnamefont
  {{Detweiler}}},\ }\bibfield  {title} {\bibinfo {title} {{Pulsar timing
  measurements and the search for gravitational waves}},\ }\href
  {https://doi.org/10.1086/157593} {\bibfield  {journal} {\bibinfo  {journal}
  {\apj}\ }\textbf {\bibinfo {volume} {234}},\ \bibinfo {pages} {1100}
  (\bibinfo {year} {1979})}\BibitemShut {NoStop}%
\bibitem [{\citenamefont {{Sazhin}}(1978)}]{1978SvA....22...36S}%
  \BibitemOpen
  \bibfield  {author} {\bibinfo {author} {\bibfnamefont {M.~V.}\ \bibnamefont
  {{Sazhin}}},\ }\bibfield  {title} {\bibinfo {title} {{Opportunities for
  detecting ultralong gravitational waves}},\ }\href@noop {} {\bibfield
  {journal} {\bibinfo  {journal} {Sov. Astron.}\ }\textbf {\bibinfo {volume}
  {22}},\ \bibinfo {pages} {36} (\bibinfo {year} {1978})}\BibitemShut {NoStop}%
\bibitem [{\citenamefont {{Mashhoon}}(1982)}]{1982MNRAS.199..659M}%
  \BibitemOpen
  \bibfield  {author} {\bibinfo {author} {\bibfnamefont {B.}~\bibnamefont
  {{Mashhoon}}},\ }\bibfield  {title} {\bibinfo {title} {{On the contribution
  of a stochastic background of gravitationnal radiation to the timing noise of
  pulsars.}},\ }\href {https://doi.org/10.1093/mnras/199.3.659} {\bibfield
  {journal} {\bibinfo  {journal} {Monthly Notices of the Royal Astronomical
  Society}\ }\textbf {\bibinfo {volume} {199}},\ \bibinfo {pages} {659}
  (\bibinfo {year} {1982})}\BibitemShut {NoStop}%
\bibitem [{\citenamefont {{Bertotti}}\ \emph {et~al.}(1983)\citenamefont
  {{Bertotti}}, \citenamefont {{Carr}},\ and\ \citenamefont
  {{Rees}}}]{1983MNRAS.203..945B}%
  \BibitemOpen
  \bibfield  {author} {\bibinfo {author} {\bibfnamefont {B.}~\bibnamefont
  {{Bertotti}}}, \bibinfo {author} {\bibfnamefont {B.~J.}\ \bibnamefont
  {{Carr}}},\ and\ \bibinfo {author} {\bibfnamefont {M.~J.}\ \bibnamefont
  {{Rees}}},\ }\bibfield  {title} {\bibinfo {title} {{Limits from the timing of
  pulsars on the cosmic gravitational wave background.}},\ }\href
  {https://doi.org/10.1093/mnras/203.4.945} {\bibfield  {journal} {\bibinfo
  {journal} {Monthly Notices of the Royal Astronomical Society}\ }\textbf
  {\bibinfo {volume} {203}},\ \bibinfo {pages} {945} (\bibinfo {year}
  {1983})}\BibitemShut {NoStop}%
\bibitem [{\citenamefont {{Hellings}}\ and\ \citenamefont
  {{Downs}}(1983)}]{1983ApJ...265L..39H}%
  \BibitemOpen
  \bibfield  {author} {\bibinfo {author} {\bibfnamefont {R.~W.}\ \bibnamefont
  {{Hellings}}}\ and\ \bibinfo {author} {\bibfnamefont {G.~S.}\ \bibnamefont
  {{Downs}}},\ }\bibfield  {title} {\bibinfo {title} {{Upper limits on the
  isotropic gravitational radiation background from pulsar timing analysis.}},\
  }\href {https://doi.org/10.1086/183954} {\bibfield  {journal} {\bibinfo
  {journal} {The Astrophysical Journal Letters}\ }\textbf {\bibinfo {volume}
  {265}},\ \bibinfo {pages} {L39} (\bibinfo {year} {1983})}\BibitemShut
  {NoStop}%
\bibitem [{\citenamefont {{Foster}}\ and\ \citenamefont
  {{Backer}}(1990)}]{1990ApJ...361..300F}%
  \BibitemOpen
  \bibfield  {author} {\bibinfo {author} {\bibfnamefont {R.~S.}\ \bibnamefont
  {{Foster}}}\ and\ \bibinfo {author} {\bibfnamefont {D.~C.}\ \bibnamefont
  {{Backer}}},\ }\bibfield  {title} {\bibinfo {title} {{Constructing a Pulsar
  Timing Array}},\ }\href {https://doi.org/10.1086/169195} {\bibfield
  {journal} {\bibinfo  {journal} {\apj}\ }\textbf {\bibinfo {volume} {361}},\
  \bibinfo {pages} {300} (\bibinfo {year} {1990})}\BibitemShut {NoStop}%
\bibitem [{\citenamefont {{Kaspi}}\ \emph {et~al.}(1994)\citenamefont
  {{Kaspi}}, \citenamefont {{Taylor}},\ and\ \citenamefont
  {{Ryba}}}]{1994ApJ...428..713K}%
  \BibitemOpen
  \bibfield  {author} {\bibinfo {author} {\bibfnamefont {V.~M.}\ \bibnamefont
  {{Kaspi}}}, \bibinfo {author} {\bibfnamefont {J.~H.}\ \bibnamefont
  {{Taylor}}},\ and\ \bibinfo {author} {\bibfnamefont {M.~F.}\ \bibnamefont
  {{Ryba}}},\ }\bibfield  {title} {\bibinfo {title} {{High-Precision Timing of
  Millisecond Pulsars. III. Long-Term Monitoring of PSRs B1855+09 and
  B1937+21}},\ }\href {https://doi.org/10.1086/174280} {\bibfield  {journal}
  {\bibinfo  {journal} {\apj}\ }\textbf {\bibinfo {volume} {428}},\ \bibinfo
  {pages} {713} (\bibinfo {year} {1994})}\BibitemShut {NoStop}%
\bibitem [{\citenamefont {{Jenet}}\ \emph {et~al.}(2005)\citenamefont
  {{Jenet}}, \citenamefont {{Hobbs}}, \citenamefont {{Lee}},\ and\
  \citenamefont {{Manchester}}}]{2005ApJ...625L.123J}%
  \BibitemOpen
  \bibfield  {author} {\bibinfo {author} {\bibfnamefont {F.~A.}\ \bibnamefont
  {{Jenet}}}, \bibinfo {author} {\bibfnamefont {G.~B.}\ \bibnamefont
  {{Hobbs}}}, \bibinfo {author} {\bibfnamefont {K.~J.}\ \bibnamefont {{Lee}}},\
  and\ \bibinfo {author} {\bibfnamefont {R.~N.}\ \bibnamefont {{Manchester}}},\
  }\bibfield  {title} {\bibinfo {title} {{Detecting the Stochastic
  Gravitational Wave Background Using Pulsar Timing}},\ }\href
  {https://doi.org/10.1086/431220} {\bibfield  {journal} {\bibinfo  {journal}
  {The Astrophysical Journal Letters}\ }\textbf {\bibinfo {volume} {625}},\
  \bibinfo {pages} {L123} (\bibinfo {year} {2005})},\ \Eprint
  {https://arxiv.org/abs/astro-ph/0504458} {arXiv:astro-ph/0504458 [astro-ph]}
  \BibitemShut {NoStop}%
\bibitem [{\citenamefont {{Jenet}}\ \emph {et~al.}(2006)\citenamefont
  {{Jenet}}, \citenamefont {{Hobbs}}, \citenamefont {{van Straten}},
  \citenamefont {{Manchester}}, \citenamefont {{Bailes}}, \citenamefont
  {{Verbiest}}, \citenamefont {{Edwards}}, \citenamefont {{Hotan}},
  \citenamefont {{Sarkissian}},\ and\ \citenamefont
  {{Ord}}}]{2006ApJ...653.1571J}%
  \BibitemOpen
  \bibfield  {author} {\bibinfo {author} {\bibfnamefont {F.~A.}\ \bibnamefont
  {{Jenet}}}, \bibinfo {author} {\bibfnamefont {G.~B.}\ \bibnamefont
  {{Hobbs}}}, \bibinfo {author} {\bibfnamefont {W.}~\bibnamefont {{van
  Straten}}}, \bibinfo {author} {\bibfnamefont {R.~N.}\ \bibnamefont
  {{Manchester}}}, \bibinfo {author} {\bibfnamefont {M.}~\bibnamefont
  {{Bailes}}}, \bibinfo {author} {\bibfnamefont {J.~P.~W.}\ \bibnamefont
  {{Verbiest}}}, \bibinfo {author} {\bibfnamefont {R.~T.}\ \bibnamefont
  {{Edwards}}}, \bibinfo {author} {\bibfnamefont {A.~W.}\ \bibnamefont
  {{Hotan}}}, \bibinfo {author} {\bibfnamefont {J.~M.}\ \bibnamefont
  {{Sarkissian}}},\ and\ \bibinfo {author} {\bibfnamefont {S.~M.}\ \bibnamefont
  {{Ord}}},\ }\bibfield  {title} {\bibinfo {title} {{Upper Bounds on the
  Low-Frequency Stochastic Gravitational Wave Background from Pulsar Timing
  Observations: Current Limits and Future Prospects}},\ }\href
  {https://doi.org/10.1086/508702} {\bibfield  {journal} {\bibinfo  {journal}
  {\apj}\ }\textbf {\bibinfo {volume} {653}},\ \bibinfo {pages} {1571}
  (\bibinfo {year} {2006})},\ \Eprint {https://arxiv.org/abs/astro-ph/0609013}
  {arXiv:astro-ph/0609013 [astro-ph]} \BibitemShut {NoStop}%
\bibitem [{\citenamefont {{Hobbs}}\ \emph {et~al.}(2009)\citenamefont
  {{Hobbs}}, \citenamefont {{Jenet}}, \citenamefont {{Lee}}, \citenamefont
  {{Verbiest}}, \citenamefont {{Yardley}}, \citenamefont {{Manchester}},
  \citenamefont {{Lommen}}, \citenamefont {{Coles}}, \citenamefont
  {{Edwards}},\ and\ \citenamefont {{Shettigara}}}]{2009MNRAS.394.1945H}%
  \BibitemOpen
  \bibfield  {author} {\bibinfo {author} {\bibfnamefont {G.}~\bibnamefont
  {{Hobbs}}}, \bibinfo {author} {\bibfnamefont {F.}~\bibnamefont {{Jenet}}},
  \bibinfo {author} {\bibfnamefont {K.~J.}\ \bibnamefont {{Lee}}}, \bibinfo
  {author} {\bibfnamefont {J.~P.~W.}\ \bibnamefont {{Verbiest}}}, \bibinfo
  {author} {\bibfnamefont {D.}~\bibnamefont {{Yardley}}}, \bibinfo {author}
  {\bibfnamefont {R.}~\bibnamefont {{Manchester}}}, \bibinfo {author}
  {\bibfnamefont {A.}~\bibnamefont {{Lommen}}}, \bibinfo {author}
  {\bibfnamefont {W.}~\bibnamefont {{Coles}}}, \bibinfo {author} {\bibfnamefont
  {R.}~\bibnamefont {{Edwards}}},\ and\ \bibinfo {author} {\bibfnamefont
  {C.}~\bibnamefont {{Shettigara}}},\ }\bibfield  {title} {\bibinfo {title}
  {{TEMPO2: a new pulsar timing package - III. Gravitational wave
  simulation}},\ }\href {https://doi.org/10.1111/j.1365-2966.2009.14391.x}
  {\bibfield  {journal} {\bibinfo  {journal} {Monthly Notices of the Royal
  Astronomical Society}\ }\textbf {\bibinfo {volume} {394}},\ \bibinfo {pages}
  {1945} (\bibinfo {year} {2009})},\ \Eprint {https://arxiv.org/abs/0901.0592}
  {arXiv:0901.0592 [astro-ph.SR]} \BibitemShut {NoStop}%
\bibitem [{\citenamefont {{Yardley}}\ \emph {et~al.}(2010)\citenamefont
  {{Yardley}}, \citenamefont {{Hobbs}}, \citenamefont {{Jenet}}, \citenamefont
  {{Verbiest}}, \citenamefont {{Wen}}, \citenamefont {{Manchester}},
  \citenamefont {{Coles}}, \citenamefont {{van Straten}}, \citenamefont
  {{Bailes}}, \citenamefont {{Bhat}}, \citenamefont {{Burke-Spolaor}},
  \citenamefont {{Champion}}, \citenamefont {{Hotan}},\ and\ \citenamefont
  {{Sarkissian}}}]{2010MNRAS.407..669Y}%
  \BibitemOpen
  \bibfield  {author} {\bibinfo {author} {\bibfnamefont {D.~R.~B.}\
  \bibnamefont {{Yardley}}}, \bibinfo {author} {\bibfnamefont {G.~B.}\
  \bibnamefont {{Hobbs}}}, \bibinfo {author} {\bibfnamefont {F.~A.}\
  \bibnamefont {{Jenet}}}, \bibinfo {author} {\bibfnamefont {J.~P.~W.}\
  \bibnamefont {{Verbiest}}}, \bibinfo {author} {\bibfnamefont {Z.~L.}\
  \bibnamefont {{Wen}}}, \bibinfo {author} {\bibfnamefont {R.~N.}\ \bibnamefont
  {{Manchester}}}, \bibinfo {author} {\bibfnamefont {W.~A.}\ \bibnamefont
  {{Coles}}}, \bibinfo {author} {\bibfnamefont {W.}~\bibnamefont {{van
  Straten}}}, \bibinfo {author} {\bibfnamefont {M.}~\bibnamefont {{Bailes}}},
  \bibinfo {author} {\bibfnamefont {N.~D.~R.}\ \bibnamefont {{Bhat}}}, \bibinfo
  {author} {\bibfnamefont {S.}~\bibnamefont {{Burke-Spolaor}}}, \bibinfo
  {author} {\bibfnamefont {D.~J.}\ \bibnamefont {{Champion}}}, \bibinfo
  {author} {\bibfnamefont {A.~W.}\ \bibnamefont {{Hotan}}},\ and\ \bibinfo
  {author} {\bibfnamefont {J.~M.}\ \bibnamefont {{Sarkissian}}},\ }\bibfield
  {title} {\bibinfo {title} {{The sensitivity of the Parkes Pulsar Timing Array
  to individual sources of gravitational waves}},\ }\href
  {https://doi.org/10.1111/j.1365-2966.2010.16949.x} {\bibfield  {journal}
  {\bibinfo  {journal} {Monthly Notices of the Royal Astronomical Society}\
  }\textbf {\bibinfo {volume} {407}},\ \bibinfo {pages} {669} (\bibinfo {year}
  {2010})},\ \Eprint {https://arxiv.org/abs/1005.1667} {arXiv:1005.1667
  [astro-ph.GA]} \BibitemShut {NoStop}%
\bibitem [{\citenamefont {{J.~P.~W. Verbiest et
  al.}}(2016)}]{2016MNRAS.458.1267V}%
  \BibitemOpen
  \bibfield  {author} {\bibinfo {author} {\bibnamefont {{J.~P.~W. Verbiest et
  al.}}},\ }\bibfield  {title} {\bibinfo {title} {{The International Pulsar
  Timing Array: First data release}},\ }\href
  {https://doi.org/10.1093/mnras/stw347} {\bibfield  {journal} {\bibinfo
  {journal} {Monthly Notices of the Royal Astronomical Society}\ }\textbf
  {\bibinfo {volume} {458}},\ \bibinfo {pages} {1267} (\bibinfo {year}
  {2016})},\ \Eprint {https://arxiv.org/abs/1602.03640} {arXiv:1602.03640
  [astro-ph.IM]} \BibitemShut {NoStop}%
\bibitem [{\citenamefont {{S. Babak et al.}}(2016)}]{2016MNRAS.455.1665B}%
  \BibitemOpen
  \bibfield  {author} {\bibinfo {author} {\bibnamefont {{S. Babak et al.}}},\
  }\bibfield  {title} {\bibinfo {title} {{European Pulsar Timing Array limits
  on continuous gravitational waves from individual supermassive black hole
  binaries}},\ }\href {https://doi.org/10.1093/mnras/stv2092} {\bibfield
  {journal} {\bibinfo  {journal} {Monthly Notices of the Royal Astronomical
  Society}\ }\textbf {\bibinfo {volume} {455}},\ \bibinfo {pages} {1665}
  (\bibinfo {year} {2016})},\ \Eprint {https://arxiv.org/abs/1509.02165}
  {arXiv:1509.02165 [astro-ph.CO]} \BibitemShut {NoStop}%
\bibitem [{\citenamefont {{Z. Arzoumanian et
  al.}}(2020)}]{2020ApJ...905L..34A}%
  \BibitemOpen
  \bibfield  {author} {\bibinfo {author} {\bibnamefont {{Z. Arzoumanian et
  al.}}},\ }\bibfield  {title} {\bibinfo {title} {{The NANOGrav 12.5 yr Data
  Set: Search for an Isotropic Stochastic Gravitational-wave Background}},\
  }\href {https://doi.org/10.3847/2041-8213/abd401} {\bibfield  {journal}
  {\bibinfo  {journal} {The Astrophysical Journal Letters}\ }\textbf {\bibinfo
  {volume} {905}},\ \bibinfo {eid} {L34} (\bibinfo {year} {2020})},\ \Eprint
  {https://arxiv.org/abs/2009.04496} {arXiv:2009.04496 [astro-ph.HE]}
  \BibitemShut {NoStop}%
\bibitem [{\citenamefont {{Ferrarese}}\ and\ \citenamefont
  {{Ford}}(2005)}]{2005SSRv..116..523F}%
  \BibitemOpen
  \bibfield  {author} {\bibinfo {author} {\bibfnamefont {L.}~\bibnamefont
  {{Ferrarese}}}\ and\ \bibinfo {author} {\bibfnamefont {H.}~\bibnamefont
  {{Ford}}},\ }\bibfield  {title} {\bibinfo {title} {{Supermassive Black Holes
  in Galactic Nuclei: Past, Present and Future Research}},\ }\href
  {https://doi.org/10.1007/s11214-005-3947-6} {\bibfield  {journal} {\bibinfo
  {journal} {Space Science Reviews}\ }\textbf {\bibinfo {volume} {116}},\
  \bibinfo {pages} {523} (\bibinfo {year} {2005})},\ \Eprint
  {https://arxiv.org/abs/astro-ph/0411247} {arXiv:astro-ph/0411247 [astro-ph]}
  \BibitemShut {NoStop}%
\bibitem [{\citenamefont {{Agazie}}\ \emph {et~al.}(2023)\citenamefont
  {{Agazie}}, \citenamefont {{Anumarlapudi}}, \citenamefont {{Archibald}},
  \citenamefont {{Arzoumanian}}, \citenamefont {{Baker}}, \citenamefont
  {{B{\'e}csy}}, \citenamefont {{Blecha}}, \citenamefont {{Brazier}},
  \citenamefont {{Brook}}, \citenamefont {{Burke-Spolaor}}, \citenamefont
  {{Burnette}}, \citenamefont {{Case}}, \citenamefont {{Charisi}},
  \citenamefont {{Chatterjee}}, \citenamefont {{Chatziioannou}}, \citenamefont
  {{Cheeseboro}}, \citenamefont {{Chen}}, \citenamefont {{Cohen}},
  \citenamefont {{Cordes}}, \citenamefont {{Cornish}}, \citenamefont
  {{Crawford}}, \citenamefont {{Cromartie}}, \citenamefont {{Crowter}},
  \citenamefont {{Cutler}}, \citenamefont {{Decesar}}, \citenamefont {{Degan}},
  \citenamefont {{Demorest}}, \citenamefont {{Deng}}, \citenamefont {{Dolch}},
  \citenamefont {{Drachler}}, \citenamefont {{Ellis}}, \citenamefont
  {{Ferrara}}, \citenamefont {{Fiore}}, \citenamefont {{Fonseca}},
  \citenamefont {{Freedman}}, \citenamefont {{Garver-Daniels}}, \citenamefont
  {{Gentile}}, \citenamefont {{Gersbach}}, \citenamefont {{Glaser}},
  \citenamefont {{Good}}, \citenamefont {{G{\"u}ltekin}}, \citenamefont
  {{Hazboun}}, \citenamefont {{Hourihane}}, \citenamefont {{Islo}},
  \citenamefont {{Jennings}}, \citenamefont {{Johnson}}, \citenamefont
  {{Jones}}, \citenamefont {{Kaiser}}, \citenamefont {{Kaplan}}, \citenamefont
  {{Kelley}}, \citenamefont {{Kerr}}, \citenamefont {{Key}}, \citenamefont
  {{Klein}}, \citenamefont {{Laal}}, \citenamefont {{Lam}}, \citenamefont
  {{Lamb}}, \citenamefont {{Lazio}}, \citenamefont {{Lewandowska}},
  \citenamefont {{Littenberg}}, \citenamefont {{Liu}}, \citenamefont
  {{Lommen}}, \citenamefont {{Lorimer}}, \citenamefont {{Luo}}, \citenamefont
  {{Lynch}}, \citenamefont {{Ma}}, \citenamefont {{Madison}}, \citenamefont
  {{Mattson}}, \citenamefont {{McEwen}}, \citenamefont {{McKee}}, \citenamefont
  {{McLaughlin}}, \citenamefont {{McMann}}, \citenamefont {{Meyers}},
  \citenamefont {{Meyers}}, \citenamefont {{Mingarelli}}, \citenamefont
  {{Mitridate}}, \citenamefont {{Natarajan}}, \citenamefont {{Ng}},
  \citenamefont {{Nice}}, \citenamefont {{Ocker}}, \citenamefont {{Olum}},
  \citenamefont {{Pennucci}}, \citenamefont {{Perera}}, \citenamefont
  {{Petrov}}, \citenamefont {{Pol}}, \citenamefont {{Radovan}}, \citenamefont
  {{Ransom}}, \citenamefont {{Ray}}, \citenamefont {{Romano}}, \citenamefont
  {{Sardesai}}, \citenamefont {{Schmiedekamp}}, \citenamefont {{Schmiedekamp}},
  \citenamefont {{Schmitz}}, \citenamefont {{Schult}}, \citenamefont
  {{Shapiro-Albert}}, \citenamefont {{Siemens}}, \citenamefont {{Simon}},
  \citenamefont {{Siwek}}, \citenamefont {{Stairs}}, \citenamefont
  {{Stinebring}}, \citenamefont {{Stovall}}, \citenamefont {{Sun}},
  \citenamefont {{Susobhanan}}, \citenamefont {{Swiggum}}, \citenamefont
  {{Taylor}}, \citenamefont {{Taylor}}, \citenamefont {{Turner}}, \citenamefont
  {{Unal}}, \citenamefont {{Vallisneri}}, \citenamefont {{van Haasteren}},
  \citenamefont {{Vigeland}}, \citenamefont {{Wahl}}, \citenamefont {{Wang}},
  \citenamefont {{Witt}}, \citenamefont {{Young}},\ and\ \citenamefont
  {{Nanograv Collaboration}}}]{2023ApJ...951L...8A}%
  \BibitemOpen
  \bibfield  {author} {\bibinfo {author} {\bibfnamefont {G.}~\bibnamefont
  {{Agazie}}}, \bibinfo {author} {\bibfnamefont {A.}~\bibnamefont
  {{Anumarlapudi}}}, \bibinfo {author} {\bibfnamefont {A.~M.}\ \bibnamefont
  {{Archibald}}}, \bibinfo {author} {\bibfnamefont {Z.}~\bibnamefont
  {{Arzoumanian}}}, \bibinfo {author} {\bibfnamefont {P.~T.}\ \bibnamefont
  {{Baker}}}, \bibinfo {author} {\bibfnamefont {B.}~\bibnamefont
  {{B{\'e}csy}}}, \bibinfo {author} {\bibfnamefont {L.}~\bibnamefont
  {{Blecha}}}, \bibinfo {author} {\bibfnamefont {A.}~\bibnamefont {{Brazier}}},
  \bibinfo {author} {\bibfnamefont {P.~R.}\ \bibnamefont {{Brook}}}, \bibinfo
  {author} {\bibfnamefont {S.}~\bibnamefont {{Burke-Spolaor}}}, \bibinfo
  {author} {\bibfnamefont {R.}~\bibnamefont {{Burnette}}}, \bibinfo {author}
  {\bibfnamefont {R.}~\bibnamefont {{Case}}}, \bibinfo {author} {\bibfnamefont
  {M.}~\bibnamefont {{Charisi}}}, \bibinfo {author} {\bibfnamefont
  {S.}~\bibnamefont {{Chatterjee}}}, \bibinfo {author} {\bibfnamefont
  {K.}~\bibnamefont {{Chatziioannou}}}, \bibinfo {author} {\bibfnamefont
  {B.~D.}\ \bibnamefont {{Cheeseboro}}}, \bibinfo {author} {\bibfnamefont
  {S.}~\bibnamefont {{Chen}}}, \bibinfo {author} {\bibfnamefont
  {T.}~\bibnamefont {{Cohen}}}, \bibinfo {author} {\bibfnamefont {J.~M.}\
  \bibnamefont {{Cordes}}}, \bibinfo {author} {\bibfnamefont {N.~J.}\
  \bibnamefont {{Cornish}}}, \bibinfo {author} {\bibfnamefont {F.}~\bibnamefont
  {{Crawford}}}, \bibinfo {author} {\bibfnamefont {H.~T.}\ \bibnamefont
  {{Cromartie}}}, \bibinfo {author} {\bibfnamefont {K.}~\bibnamefont
  {{Crowter}}}, \bibinfo {author} {\bibfnamefont {C.~J.}\ \bibnamefont
  {{Cutler}}}, \bibinfo {author} {\bibfnamefont {M.~E.}\ \bibnamefont
  {{Decesar}}}, \bibinfo {author} {\bibfnamefont {D.}~\bibnamefont {{Degan}}},
  \bibinfo {author} {\bibfnamefont {P.~B.}\ \bibnamefont {{Demorest}}},
  \bibinfo {author} {\bibfnamefont {H.}~\bibnamefont {{Deng}}}, \bibinfo
  {author} {\bibfnamefont {T.}~\bibnamefont {{Dolch}}}, \bibinfo {author}
  {\bibfnamefont {B.}~\bibnamefont {{Drachler}}}, \bibinfo {author}
  {\bibfnamefont {J.~A.}\ \bibnamefont {{Ellis}}}, \bibinfo {author}
  {\bibfnamefont {E.~C.}\ \bibnamefont {{Ferrara}}}, \bibinfo {author}
  {\bibfnamefont {W.}~\bibnamefont {{Fiore}}}, \bibinfo {author} {\bibfnamefont
  {E.}~\bibnamefont {{Fonseca}}}, \bibinfo {author} {\bibfnamefont {G.~E.}\
  \bibnamefont {{Freedman}}}, \bibinfo {author} {\bibfnamefont
  {N.}~\bibnamefont {{Garver-Daniels}}}, \bibinfo {author} {\bibfnamefont
  {P.~A.}\ \bibnamefont {{Gentile}}}, \bibinfo {author} {\bibfnamefont {K.~A.}\
  \bibnamefont {{Gersbach}}}, \bibinfo {author} {\bibfnamefont
  {J.}~\bibnamefont {{Glaser}}}, \bibinfo {author} {\bibfnamefont {D.~C.}\
  \bibnamefont {{Good}}}, \bibinfo {author} {\bibfnamefont {K.}~\bibnamefont
  {{G{\"u}ltekin}}}, \bibinfo {author} {\bibfnamefont {J.~S.}\ \bibnamefont
  {{Hazboun}}}, \bibinfo {author} {\bibfnamefont {S.}~\bibnamefont
  {{Hourihane}}}, \bibinfo {author} {\bibfnamefont {K.}~\bibnamefont {{Islo}}},
  \bibinfo {author} {\bibfnamefont {R.~J.}\ \bibnamefont {{Jennings}}},
  \bibinfo {author} {\bibfnamefont {A.~D.}\ \bibnamefont {{Johnson}}}, \bibinfo
  {author} {\bibfnamefont {M.~L.}\ \bibnamefont {{Jones}}}, \bibinfo {author}
  {\bibfnamefont {A.~R.}\ \bibnamefont {{Kaiser}}}, \bibinfo {author}
  {\bibfnamefont {D.~L.}\ \bibnamefont {{Kaplan}}}, \bibinfo {author}
  {\bibfnamefont {L.~Z.}\ \bibnamefont {{Kelley}}}, \bibinfo {author}
  {\bibfnamefont {M.}~\bibnamefont {{Kerr}}}, \bibinfo {author} {\bibfnamefont
  {J.~S.}\ \bibnamefont {{Key}}}, \bibinfo {author} {\bibfnamefont {T.~C.}\
  \bibnamefont {{Klein}}}, \bibinfo {author} {\bibfnamefont {N.}~\bibnamefont
  {{Laal}}}, \bibinfo {author} {\bibfnamefont {M.~T.}\ \bibnamefont {{Lam}}},
  \bibinfo {author} {\bibfnamefont {W.~G.}\ \bibnamefont {{Lamb}}}, \bibinfo
  {author} {\bibfnamefont {T.~J.~W.}\ \bibnamefont {{Lazio}}}, \bibinfo
  {author} {\bibfnamefont {N.}~\bibnamefont {{Lewandowska}}}, \bibinfo {author}
  {\bibfnamefont {T.~B.}\ \bibnamefont {{Littenberg}}}, \bibinfo {author}
  {\bibfnamefont {T.}~\bibnamefont {{Liu}}}, \bibinfo {author} {\bibfnamefont
  {A.}~\bibnamefont {{Lommen}}}, \bibinfo {author} {\bibfnamefont {D.~R.}\
  \bibnamefont {{Lorimer}}}, \bibinfo {author} {\bibfnamefont {J.}~\bibnamefont
  {{Luo}}}, \bibinfo {author} {\bibfnamefont {R.~S.}\ \bibnamefont {{Lynch}}},
  \bibinfo {author} {\bibfnamefont {C.-P.}\ \bibnamefont {{Ma}}}, \bibinfo
  {author} {\bibfnamefont {D.~R.}\ \bibnamefont {{Madison}}}, \bibinfo {author}
  {\bibfnamefont {M.~A.}\ \bibnamefont {{Mattson}}}, \bibinfo {author}
  {\bibfnamefont {A.}~\bibnamefont {{McEwen}}}, \bibinfo {author}
  {\bibfnamefont {J.~W.}\ \bibnamefont {{McKee}}}, \bibinfo {author}
  {\bibfnamefont {M.~A.}\ \bibnamefont {{McLaughlin}}}, \bibinfo {author}
  {\bibfnamefont {N.}~\bibnamefont {{McMann}}}, \bibinfo {author}
  {\bibfnamefont {B.~W.}\ \bibnamefont {{Meyers}}}, \bibinfo {author}
  {\bibfnamefont {P.~M.}\ \bibnamefont {{Meyers}}}, \bibinfo {author}
  {\bibfnamefont {C.~M.~F.}\ \bibnamefont {{Mingarelli}}}, \bibinfo {author}
  {\bibfnamefont {A.}~\bibnamefont {{Mitridate}}}, \bibinfo {author}
  {\bibfnamefont {P.}~\bibnamefont {{Natarajan}}}, \bibinfo {author}
  {\bibfnamefont {C.}~\bibnamefont {{Ng}}}, \bibinfo {author} {\bibfnamefont
  {D.~J.}\ \bibnamefont {{Nice}}}, \bibinfo {author} {\bibfnamefont {S.~K.}\
  \bibnamefont {{Ocker}}}, \bibinfo {author} {\bibfnamefont {K.~D.}\
  \bibnamefont {{Olum}}}, \bibinfo {author} {\bibfnamefont {T.~T.}\
  \bibnamefont {{Pennucci}}}, \bibinfo {author} {\bibfnamefont {B.~B.~P.}\
  \bibnamefont {{Perera}}}, \bibinfo {author} {\bibfnamefont {P.}~\bibnamefont
  {{Petrov}}}, \bibinfo {author} {\bibfnamefont {N.~S.}\ \bibnamefont {{Pol}}},
  \bibinfo {author} {\bibfnamefont {H.~A.}\ \bibnamefont {{Radovan}}}, \bibinfo
  {author} {\bibfnamefont {S.~M.}\ \bibnamefont {{Ransom}}}, \bibinfo {author}
  {\bibfnamefont {P.~S.}\ \bibnamefont {{Ray}}}, \bibinfo {author}
  {\bibfnamefont {J.~D.}\ \bibnamefont {{Romano}}}, \bibinfo {author}
  {\bibfnamefont {S.~C.}\ \bibnamefont {{Sardesai}}}, \bibinfo {author}
  {\bibfnamefont {A.}~\bibnamefont {{Schmiedekamp}}}, \bibinfo {author}
  {\bibfnamefont {C.}~\bibnamefont {{Schmiedekamp}}}, \bibinfo {author}
  {\bibfnamefont {K.}~\bibnamefont {{Schmitz}}}, \bibinfo {author}
  {\bibfnamefont {L.}~\bibnamefont {{Schult}}}, \bibinfo {author}
  {\bibfnamefont {B.~J.}\ \bibnamefont {{Shapiro-Albert}}}, \bibinfo {author}
  {\bibfnamefont {X.}~\bibnamefont {{Siemens}}}, \bibinfo {author}
  {\bibfnamefont {J.}~\bibnamefont {{Simon}}}, \bibinfo {author} {\bibfnamefont
  {M.~S.}\ \bibnamefont {{Siwek}}}, \bibinfo {author} {\bibfnamefont {I.~H.}\
  \bibnamefont {{Stairs}}}, \bibinfo {author} {\bibfnamefont {D.~R.}\
  \bibnamefont {{Stinebring}}}, \bibinfo {author} {\bibfnamefont
  {K.}~\bibnamefont {{Stovall}}}, \bibinfo {author} {\bibfnamefont {J.~P.}\
  \bibnamefont {{Sun}}}, \bibinfo {author} {\bibfnamefont {A.}~\bibnamefont
  {{Susobhanan}}}, \bibinfo {author} {\bibfnamefont {J.~K.}\ \bibnamefont
  {{Swiggum}}}, \bibinfo {author} {\bibfnamefont {J.}~\bibnamefont {{Taylor}}},
  \bibinfo {author} {\bibfnamefont {S.~R.}\ \bibnamefont {{Taylor}}}, \bibinfo
  {author} {\bibfnamefont {J.~E.}\ \bibnamefont {{Turner}}}, \bibinfo {author}
  {\bibfnamefont {C.}~\bibnamefont {{Unal}}}, \bibinfo {author} {\bibfnamefont
  {M.}~\bibnamefont {{Vallisneri}}}, \bibinfo {author} {\bibfnamefont
  {R.}~\bibnamefont {{van Haasteren}}}, \bibinfo {author} {\bibfnamefont
  {S.~J.}\ \bibnamefont {{Vigeland}}}, \bibinfo {author} {\bibfnamefont
  {H.~M.}\ \bibnamefont {{Wahl}}}, \bibinfo {author} {\bibfnamefont
  {Q.}~\bibnamefont {{Wang}}}, \bibinfo {author} {\bibfnamefont {C.~A.}\
  \bibnamefont {{Witt}}}, \bibinfo {author} {\bibfnamefont {O.}~\bibnamefont
  {{Young}}},\ and\ \bibinfo {author} {\bibnamefont {{Nanograv
  Collaboration}}},\ }\bibfield  {title} {\bibinfo {title} {{The NANOGrav 15 yr
  Data Set: Evidence for a Gravitational-wave Background}},\ }\href
  {https://doi.org/10.3847/2041-8213/acdac6} {\bibfield  {journal} {\bibinfo
  {journal} {The Astrophysical Journal Letters}\ }\textbf {\bibinfo {volume}
  {951}},\ \bibinfo {eid} {L8} (\bibinfo {year} {2023})},\ \Eprint
  {https://arxiv.org/abs/2306.16213} {arXiv:2306.16213 [astro-ph.HE]}
  \BibitemShut {NoStop}%
\bibitem [{\citenamefont {{EPTA Collaboration}}\ \emph
  {et~al.}(2023)\citenamefont {{EPTA Collaboration}}, \citenamefont {{InPTA
  Collaboration}}, \citenamefont {{Antoniadis}}, \citenamefont {{Arumugam}},
  \citenamefont {{Arumugam}}, \citenamefont {{Babak}}, \citenamefont
  {{Bagchi}}, \citenamefont {{Bak Nielsen}}, \citenamefont {{Bassa}},
  \citenamefont {{Bathula}}, \citenamefont {{Berthereau}}, \citenamefont
  {{Bonetti}}, \citenamefont {{Bortolas}}, \citenamefont {{Brook}},
  \citenamefont {{Burgay}}, \citenamefont {{Caballero}}, \citenamefont
  {{Chalumeau}}, \citenamefont {{Champion}}, \citenamefont {{Chanlaridis}},
  \citenamefont {{Chen}}, \citenamefont {{Cognard}}, \citenamefont
  {{Dandapat}}, \citenamefont {{Deb}}, \citenamefont {{Desai}}, \citenamefont
  {{Desvignes}}, \citenamefont {{Dhanda-Batra}}, \citenamefont {{Dwivedi}},
  \citenamefont {{Falxa}}, \citenamefont {{Ferdman}}, \citenamefont
  {{Franchini}}, \citenamefont {{Gair}}, \citenamefont {{Goncharov}},
  \citenamefont {{Gopakumar}}, \citenamefont {{Graikou}}, \citenamefont
  {{Grie{\ss}meier}}, \citenamefont {{Guillemot}}, \citenamefont {{Guo}},
  \citenamefont {{Gupta}}, \citenamefont {{Hisano}}, \citenamefont {{Hu}},
  \citenamefont {{Iraci}}, \citenamefont {{Izquierdo-Villalba}}, \citenamefont
  {{Jang}}, \citenamefont {{Jawor}}, \citenamefont {{Janssen}}, \citenamefont
  {{Jessner}}, \citenamefont {{Joshi}}, \citenamefont {{Kareem}}, \citenamefont
  {{Karuppusamy}}, \citenamefont {{Keane}}, \citenamefont {{Keith}},
  \citenamefont {{Kharbanda}}, \citenamefont {{Kikunaga}}, \citenamefont
  {{Kolhe}}, \citenamefont {{Kramer}}, \citenamefont {{Krishnakumar}},
  \citenamefont {{Lackeos}}, \citenamefont {{Lee}}, \citenamefont {{Liu}},
  \citenamefont {{Liu}}, \citenamefont {{Lyne}}, \citenamefont {{McKee}},
  \citenamefont {{Maan}}, \citenamefont {{Main}}, \citenamefont {{Mickaliger}},
  \citenamefont {{Ni{\c{t}}u}}, \citenamefont {{Nobleson}}, \citenamefont
  {{Paladi}}, \citenamefont {{Parthasarathy}}, \citenamefont {{Perera}},
  \citenamefont {{Perrodin}}, \citenamefont {{Petiteau}}, \citenamefont
  {{Porayko}}, \citenamefont {{Possenti}}, \citenamefont {{Prabu}},
  \citenamefont {{Quelquejay Leclere}}, \citenamefont {{Rana}}, \citenamefont
  {{Samajdar}}, \citenamefont {{Sanidas}}, \citenamefont {{Sesana}},
  \citenamefont {{Shaifullah}}, \citenamefont {{Singha}}, \citenamefont
  {{Speri}}, \citenamefont {{Spiewak}}, \citenamefont {{Srivastava}},
  \citenamefont {{Stappers}}, \citenamefont {{Surnis}}, \citenamefont
  {{Susarla}}, \citenamefont {{Susobhanan}}, \citenamefont {{Takahashi}},
  \citenamefont {{Tarafdar}}, \citenamefont {{Theureau}}, \citenamefont
  {{Tiburzi}}, \citenamefont {{van der Wateren}}, \citenamefont {{Vecchio}},
  \citenamefont {{Venkatraman Krishnan}}, \citenamefont {{Verbiest}},
  \citenamefont {{Wang}}, \citenamefont {{Wang}},\ and\ \citenamefont
  {{Wu}}}]{2023A&A...678A..50E}%
  \BibitemOpen
  \bibfield  {author} {\bibinfo {author} {\bibnamefont {{EPTA Collaboration}}},
  \bibinfo {author} {\bibnamefont {{InPTA Collaboration}}}, \bibinfo {author}
  {\bibfnamefont {J.}~\bibnamefont {{Antoniadis}}}, \bibinfo {author}
  {\bibfnamefont {P.}~\bibnamefont {{Arumugam}}}, \bibinfo {author}
  {\bibfnamefont {S.}~\bibnamefont {{Arumugam}}}, \bibinfo {author}
  {\bibfnamefont {S.}~\bibnamefont {{Babak}}}, \bibinfo {author} {\bibfnamefont
  {M.}~\bibnamefont {{Bagchi}}}, \bibinfo {author} {\bibfnamefont {A.~S.}\
  \bibnamefont {{Bak Nielsen}}}, \bibinfo {author} {\bibfnamefont {C.~G.}\
  \bibnamefont {{Bassa}}}, \bibinfo {author} {\bibfnamefont {A.}~\bibnamefont
  {{Bathula}}}, \bibinfo {author} {\bibfnamefont {A.}~\bibnamefont
  {{Berthereau}}}, \bibinfo {author} {\bibfnamefont {M.}~\bibnamefont
  {{Bonetti}}}, \bibinfo {author} {\bibfnamefont {E.}~\bibnamefont
  {{Bortolas}}}, \bibinfo {author} {\bibfnamefont {P.~R.}\ \bibnamefont
  {{Brook}}}, \bibinfo {author} {\bibfnamefont {M.}~\bibnamefont {{Burgay}}},
  \bibinfo {author} {\bibfnamefont {R.~N.}\ \bibnamefont {{Caballero}}},
  \bibinfo {author} {\bibfnamefont {A.}~\bibnamefont {{Chalumeau}}}, \bibinfo
  {author} {\bibfnamefont {D.~J.}\ \bibnamefont {{Champion}}}, \bibinfo
  {author} {\bibfnamefont {S.}~\bibnamefont {{Chanlaridis}}}, \bibinfo {author}
  {\bibfnamefont {S.}~\bibnamefont {{Chen}}}, \bibinfo {author} {\bibfnamefont
  {I.}~\bibnamefont {{Cognard}}}, \bibinfo {author} {\bibfnamefont
  {S.}~\bibnamefont {{Dandapat}}}, \bibinfo {author} {\bibfnamefont
  {D.}~\bibnamefont {{Deb}}}, \bibinfo {author} {\bibfnamefont
  {S.}~\bibnamefont {{Desai}}}, \bibinfo {author} {\bibfnamefont
  {G.}~\bibnamefont {{Desvignes}}}, \bibinfo {author} {\bibfnamefont
  {N.}~\bibnamefont {{Dhanda-Batra}}}, \bibinfo {author} {\bibfnamefont
  {C.}~\bibnamefont {{Dwivedi}}}, \bibinfo {author} {\bibfnamefont
  {M.}~\bibnamefont {{Falxa}}}, \bibinfo {author} {\bibfnamefont {R.~D.}\
  \bibnamefont {{Ferdman}}}, \bibinfo {author} {\bibfnamefont {A.}~\bibnamefont
  {{Franchini}}}, \bibinfo {author} {\bibfnamefont {J.~R.}\ \bibnamefont
  {{Gair}}}, \bibinfo {author} {\bibfnamefont {B.}~\bibnamefont {{Goncharov}}},
  \bibinfo {author} {\bibfnamefont {A.}~\bibnamefont {{Gopakumar}}}, \bibinfo
  {author} {\bibfnamefont {E.}~\bibnamefont {{Graikou}}}, \bibinfo {author}
  {\bibfnamefont {J.~M.}\ \bibnamefont {{Grie{\ss}meier}}}, \bibinfo {author}
  {\bibfnamefont {L.}~\bibnamefont {{Guillemot}}}, \bibinfo {author}
  {\bibfnamefont {Y.~J.}\ \bibnamefont {{Guo}}}, \bibinfo {author}
  {\bibfnamefont {Y.}~\bibnamefont {{Gupta}}}, \bibinfo {author} {\bibfnamefont
  {S.}~\bibnamefont {{Hisano}}}, \bibinfo {author} {\bibfnamefont
  {H.}~\bibnamefont {{Hu}}}, \bibinfo {author} {\bibfnamefont {F.}~\bibnamefont
  {{Iraci}}}, \bibinfo {author} {\bibfnamefont {D.}~\bibnamefont
  {{Izquierdo-Villalba}}}, \bibinfo {author} {\bibfnamefont {J.}~\bibnamefont
  {{Jang}}}, \bibinfo {author} {\bibfnamefont {J.}~\bibnamefont {{Jawor}}},
  \bibinfo {author} {\bibfnamefont {G.~H.}\ \bibnamefont {{Janssen}}}, \bibinfo
  {author} {\bibfnamefont {A.}~\bibnamefont {{Jessner}}}, \bibinfo {author}
  {\bibfnamefont {B.~C.}\ \bibnamefont {{Joshi}}}, \bibinfo {author}
  {\bibfnamefont {F.}~\bibnamefont {{Kareem}}}, \bibinfo {author}
  {\bibfnamefont {R.}~\bibnamefont {{Karuppusamy}}}, \bibinfo {author}
  {\bibfnamefont {E.~F.}\ \bibnamefont {{Keane}}}, \bibinfo {author}
  {\bibfnamefont {M.~J.}\ \bibnamefont {{Keith}}}, \bibinfo {author}
  {\bibfnamefont {D.}~\bibnamefont {{Kharbanda}}}, \bibinfo {author}
  {\bibfnamefont {T.}~\bibnamefont {{Kikunaga}}}, \bibinfo {author}
  {\bibfnamefont {N.}~\bibnamefont {{Kolhe}}}, \bibinfo {author} {\bibfnamefont
  {M.}~\bibnamefont {{Kramer}}}, \bibinfo {author} {\bibfnamefont {M.~A.}\
  \bibnamefont {{Krishnakumar}}}, \bibinfo {author} {\bibfnamefont
  {K.}~\bibnamefont {{Lackeos}}}, \bibinfo {author} {\bibfnamefont {K.~J.}\
  \bibnamefont {{Lee}}}, \bibinfo {author} {\bibfnamefont {K.}~\bibnamefont
  {{Liu}}}, \bibinfo {author} {\bibfnamefont {Y.}~\bibnamefont {{Liu}}},
  \bibinfo {author} {\bibfnamefont {A.~G.}\ \bibnamefont {{Lyne}}}, \bibinfo
  {author} {\bibfnamefont {J.~W.}\ \bibnamefont {{McKee}}}, \bibinfo {author}
  {\bibfnamefont {Y.}~\bibnamefont {{Maan}}}, \bibinfo {author} {\bibfnamefont
  {R.~A.}\ \bibnamefont {{Main}}}, \bibinfo {author} {\bibfnamefont {M.~B.}\
  \bibnamefont {{Mickaliger}}}, \bibinfo {author} {\bibfnamefont {I.~C.}\
  \bibnamefont {{Ni{\c{t}}u}}}, \bibinfo {author} {\bibfnamefont
  {K.}~\bibnamefont {{Nobleson}}}, \bibinfo {author} {\bibfnamefont {A.~K.}\
  \bibnamefont {{Paladi}}}, \bibinfo {author} {\bibfnamefont {A.}~\bibnamefont
  {{Parthasarathy}}}, \bibinfo {author} {\bibfnamefont {B.~B.~P.}\ \bibnamefont
  {{Perera}}}, \bibinfo {author} {\bibfnamefont {D.}~\bibnamefont
  {{Perrodin}}}, \bibinfo {author} {\bibfnamefont {A.}~\bibnamefont
  {{Petiteau}}}, \bibinfo {author} {\bibfnamefont {N.~K.}\ \bibnamefont
  {{Porayko}}}, \bibinfo {author} {\bibfnamefont {A.}~\bibnamefont
  {{Possenti}}}, \bibinfo {author} {\bibfnamefont {T.}~\bibnamefont {{Prabu}}},
  \bibinfo {author} {\bibfnamefont {H.}~\bibnamefont {{Quelquejay Leclere}}},
  \bibinfo {author} {\bibfnamefont {P.}~\bibnamefont {{Rana}}}, \bibinfo
  {author} {\bibfnamefont {A.}~\bibnamefont {{Samajdar}}}, \bibinfo {author}
  {\bibfnamefont {S.~A.}\ \bibnamefont {{Sanidas}}}, \bibinfo {author}
  {\bibfnamefont {A.}~\bibnamefont {{Sesana}}}, \bibinfo {author}
  {\bibfnamefont {G.}~\bibnamefont {{Shaifullah}}}, \bibinfo {author}
  {\bibfnamefont {J.}~\bibnamefont {{Singha}}}, \bibinfo {author}
  {\bibfnamefont {L.}~\bibnamefont {{Speri}}}, \bibinfo {author} {\bibfnamefont
  {R.}~\bibnamefont {{Spiewak}}}, \bibinfo {author} {\bibfnamefont
  {A.}~\bibnamefont {{Srivastava}}}, \bibinfo {author} {\bibfnamefont {B.~W.}\
  \bibnamefont {{Stappers}}}, \bibinfo {author} {\bibfnamefont
  {M.}~\bibnamefont {{Surnis}}}, \bibinfo {author} {\bibfnamefont {S.~C.}\
  \bibnamefont {{Susarla}}}, \bibinfo {author} {\bibfnamefont {A.}~\bibnamefont
  {{Susobhanan}}}, \bibinfo {author} {\bibfnamefont {K.}~\bibnamefont
  {{Takahashi}}}, \bibinfo {author} {\bibfnamefont {P.}~\bibnamefont
  {{Tarafdar}}}, \bibinfo {author} {\bibfnamefont {G.}~\bibnamefont
  {{Theureau}}}, \bibinfo {author} {\bibfnamefont {C.}~\bibnamefont
  {{Tiburzi}}}, \bibinfo {author} {\bibfnamefont {E.}~\bibnamefont {{van der
  Wateren}}}, \bibinfo {author} {\bibfnamefont {A.}~\bibnamefont {{Vecchio}}},
  \bibinfo {author} {\bibfnamefont {V.}~\bibnamefont {{Venkatraman Krishnan}}},
  \bibinfo {author} {\bibfnamefont {J.~P.~W.}\ \bibnamefont {{Verbiest}}},
  \bibinfo {author} {\bibfnamefont {J.}~\bibnamefont {{Wang}}}, \bibinfo
  {author} {\bibfnamefont {L.}~\bibnamefont {{Wang}}},\ and\ \bibinfo {author}
  {\bibfnamefont {Z.}~\bibnamefont {{Wu}}},\ }\bibfield  {title} {\bibinfo
  {title} {{The second data release from the European Pulsar Timing Array. III.
  Search for gravitational wave signals}},\ }\href
  {https://doi.org/10.1051/0004-6361/202346844} {\bibfield  {journal} {\bibinfo
   {journal} {Astronomy \& Astrophysics}\ }\textbf {\bibinfo {volume} {678}},\
  \bibinfo {eid} {A50} (\bibinfo {year} {2023})},\ \Eprint
  {https://arxiv.org/abs/2306.16214} {arXiv:2306.16214 [astro-ph.HE]}
  \BibitemShut {NoStop}%
\bibitem [{\citenamefont {{Reardon}}\ \emph {et~al.}(2023)\citenamefont
  {{Reardon}}, \citenamefont {{Zic}}, \citenamefont {{Shannon}}, \citenamefont
  {{Hobbs}}, \citenamefont {{Bailes}}, \citenamefont {{Di Marco}},
  \citenamefont {{Kapur}}, \citenamefont {{Rogers}}, \citenamefont {{Thrane}},
  \citenamefont {{Askew}}, \citenamefont {{Bhat}}, \citenamefont {{Cameron}},
  \citenamefont {{Cury{\l}o}}, \citenamefont {{Coles}}, \citenamefont {{Dai}},
  \citenamefont {{Goncharov}}, \citenamefont {{Kerr}}, \citenamefont
  {{Kulkarni}}, \citenamefont {{Levin}}, \citenamefont {{Lower}}, \citenamefont
  {{Manchester}}, \citenamefont {{Mandow}}, \citenamefont {{Miles}},
  \citenamefont {{Nathan}}, \citenamefont {{Os{\l}owski}}, \citenamefont
  {{Russell}}, \citenamefont {{Spiewak}}, \citenamefont {{Zhang}},\ and\
  \citenamefont {{Zhu}}}]{2023ApJ...951L...6R}%
  \BibitemOpen
  \bibfield  {author} {\bibinfo {author} {\bibfnamefont {D.~J.}\ \bibnamefont
  {{Reardon}}}, \bibinfo {author} {\bibfnamefont {A.}~\bibnamefont {{Zic}}},
  \bibinfo {author} {\bibfnamefont {R.~M.}\ \bibnamefont {{Shannon}}}, \bibinfo
  {author} {\bibfnamefont {G.~B.}\ \bibnamefont {{Hobbs}}}, \bibinfo {author}
  {\bibfnamefont {M.}~\bibnamefont {{Bailes}}}, \bibinfo {author}
  {\bibfnamefont {V.}~\bibnamefont {{Di Marco}}}, \bibinfo {author}
  {\bibfnamefont {A.}~\bibnamefont {{Kapur}}}, \bibinfo {author} {\bibfnamefont
  {A.~F.}\ \bibnamefont {{Rogers}}}, \bibinfo {author} {\bibfnamefont
  {E.}~\bibnamefont {{Thrane}}}, \bibinfo {author} {\bibfnamefont
  {J.}~\bibnamefont {{Askew}}}, \bibinfo {author} {\bibfnamefont {N.~D.~R.}\
  \bibnamefont {{Bhat}}}, \bibinfo {author} {\bibfnamefont {A.}~\bibnamefont
  {{Cameron}}}, \bibinfo {author} {\bibfnamefont {M.}~\bibnamefont
  {{Cury{\l}o}}}, \bibinfo {author} {\bibfnamefont {W.~A.}\ \bibnamefont
  {{Coles}}}, \bibinfo {author} {\bibfnamefont {S.}~\bibnamefont {{Dai}}},
  \bibinfo {author} {\bibfnamefont {B.}~\bibnamefont {{Goncharov}}}, \bibinfo
  {author} {\bibfnamefont {M.}~\bibnamefont {{Kerr}}}, \bibinfo {author}
  {\bibfnamefont {A.}~\bibnamefont {{Kulkarni}}}, \bibinfo {author}
  {\bibfnamefont {Y.}~\bibnamefont {{Levin}}}, \bibinfo {author} {\bibfnamefont
  {M.~E.}\ \bibnamefont {{Lower}}}, \bibinfo {author} {\bibfnamefont {R.~N.}\
  \bibnamefont {{Manchester}}}, \bibinfo {author} {\bibfnamefont
  {R.}~\bibnamefont {{Mandow}}}, \bibinfo {author} {\bibfnamefont {M.~T.}\
  \bibnamefont {{Miles}}}, \bibinfo {author} {\bibfnamefont {R.~S.}\
  \bibnamefont {{Nathan}}}, \bibinfo {author} {\bibfnamefont {S.}~\bibnamefont
  {{Os{\l}owski}}}, \bibinfo {author} {\bibfnamefont {C.~J.}\ \bibnamefont
  {{Russell}}}, \bibinfo {author} {\bibfnamefont {R.}~\bibnamefont
  {{Spiewak}}}, \bibinfo {author} {\bibfnamefont {S.}~\bibnamefont {{Zhang}}},\
  and\ \bibinfo {author} {\bibfnamefont {X.-J.}\ \bibnamefont {{Zhu}}},\
  }\bibfield  {title} {\bibinfo {title} {{Search for an Isotropic
  Gravitational-wave Background with the Parkes Pulsar Timing Array}},\ }\href
  {https://doi.org/10.3847/2041-8213/acdd02} {\bibfield  {journal} {\bibinfo
  {journal} {The Astrophysical Journal Letters}\ }\textbf {\bibinfo {volume}
  {951}},\ \bibinfo {eid} {L6} (\bibinfo {year} {2023})},\ \Eprint
  {https://arxiv.org/abs/2306.16215} {arXiv:2306.16215 [astro-ph.HE]}
  \BibitemShut {NoStop}%
\bibitem [{\citenamefont {Xu}\ \emph {et~al.}(2023)\citenamefont {Xu},
  \citenamefont {Chen}, \citenamefont {Guo}, \citenamefont {Jiang},
  \citenamefont {Wang}, \citenamefont {Xu}, \citenamefont {Xue}, \citenamefont
  {Caballero}, \citenamefont {Yuan}, \citenamefont {Xu}, \citenamefont {Wang},
  \citenamefont {Hao}, \citenamefont {Luo}, \citenamefont {Lee}, \citenamefont
  {Han}, \citenamefont {Jiang}, \citenamefont {Shen}, \citenamefont {Wang},
  \citenamefont {Wang}, \citenamefont {Xu}, \citenamefont {Wu}, \citenamefont
  {Manchester}, \citenamefont {Qian}, \citenamefont {Guan}, \citenamefont
  {Huang}, \citenamefont {Sun},\ and\ \citenamefont {Zhu}}]{Xu_2023}%
  \BibitemOpen
  \bibfield  {author} {\bibinfo {author} {\bibfnamefont {H.}~\bibnamefont
  {Xu}}, \bibinfo {author} {\bibfnamefont {S.}~\bibnamefont {Chen}}, \bibinfo
  {author} {\bibfnamefont {Y.}~\bibnamefont {Guo}}, \bibinfo {author}
  {\bibfnamefont {J.}~\bibnamefont {Jiang}}, \bibinfo {author} {\bibfnamefont
  {B.}~\bibnamefont {Wang}}, \bibinfo {author} {\bibfnamefont {J.}~\bibnamefont
  {Xu}}, \bibinfo {author} {\bibfnamefont {Z.}~\bibnamefont {Xue}}, \bibinfo
  {author} {\bibfnamefont {R.~N.}\ \bibnamefont {Caballero}}, \bibinfo {author}
  {\bibfnamefont {J.}~\bibnamefont {Yuan}}, \bibinfo {author} {\bibfnamefont
  {Y.}~\bibnamefont {Xu}}, \bibinfo {author} {\bibfnamefont {J.}~\bibnamefont
  {Wang}}, \bibinfo {author} {\bibfnamefont {L.}~\bibnamefont {Hao}}, \bibinfo
  {author} {\bibfnamefont {J.}~\bibnamefont {Luo}}, \bibinfo {author}
  {\bibfnamefont {K.}~\bibnamefont {Lee}}, \bibinfo {author} {\bibfnamefont
  {J.}~\bibnamefont {Han}}, \bibinfo {author} {\bibfnamefont {P.}~\bibnamefont
  {Jiang}}, \bibinfo {author} {\bibfnamefont {Z.}~\bibnamefont {Shen}},
  \bibinfo {author} {\bibfnamefont {M.}~\bibnamefont {Wang}}, \bibinfo {author}
  {\bibfnamefont {N.}~\bibnamefont {Wang}}, \bibinfo {author} {\bibfnamefont
  {R.}~\bibnamefont {Xu}}, \bibinfo {author} {\bibfnamefont {X.}~\bibnamefont
  {Wu}}, \bibinfo {author} {\bibfnamefont {R.}~\bibnamefont {Manchester}},
  \bibinfo {author} {\bibfnamefont {L.}~\bibnamefont {Qian}}, \bibinfo {author}
  {\bibfnamefont {X.}~\bibnamefont {Guan}}, \bibinfo {author} {\bibfnamefont
  {M.}~\bibnamefont {Huang}}, \bibinfo {author} {\bibfnamefont
  {C.}~\bibnamefont {Sun}},\ and\ \bibinfo {author} {\bibfnamefont
  {Y.}~\bibnamefont {Zhu}},\ }\bibfield  {title} {\bibinfo {title} {Searching
  for the nano-hertz stochastic gravitational wave background with the chinese
  pulsar timing array data release i},\ }\href
  {https://doi.org/10.1088/1674-4527/acdfa5} {\bibfield  {journal} {\bibinfo
  {journal} {Research in Astronomy and Astrophysics}\ }\textbf {\bibinfo
  {volume} {23}},\ \bibinfo {pages} {075024} (\bibinfo {year}
  {2023})}\BibitemShut {NoStop}%
\bibitem [{\citenamefont {{Bustamante-Rosell}}\ \emph
  {et~al.}(2022)\citenamefont {{Bustamante-Rosell}}, \citenamefont {{Meyers}},
  \citenamefont {{Pearson}}, \citenamefont {{Trendafilova}},\ and\
  \citenamefont {{Zimmerman}}}]{2022PhRvD.105d4005B}%
  \BibitemOpen
  \bibfield  {author} {\bibinfo {author} {\bibfnamefont {M.~J.}\ \bibnamefont
  {{Bustamante-Rosell}}}, \bibinfo {author} {\bibfnamefont {J.}~\bibnamefont
  {{Meyers}}}, \bibinfo {author} {\bibfnamefont {N.}~\bibnamefont {{Pearson}}},
  \bibinfo {author} {\bibfnamefont {C.}~\bibnamefont {{Trendafilova}}},\ and\
  \bibinfo {author} {\bibfnamefont {A.}~\bibnamefont {{Zimmerman}}},\
  }\bibfield  {title} {\bibinfo {title} {{Gravitational wave timing array}},\
  }\href {https://doi.org/10.1103/PhysRevD.105.044005} {\bibfield  {journal}
  {\bibinfo  {journal} {Physical Review D}\ }\textbf {\bibinfo {volume}
  {105}},\ \bibinfo {eid} {044005} (\bibinfo {year} {2022})},\ \Eprint
  {https://arxiv.org/abs/2107.02788} {arXiv:2107.02788 [gr-qc]} \BibitemShut
  {NoStop}%
\bibitem [{\citenamefont {{S. Kawamura et al.}}(2011)}]{DECIGO}%
  \BibitemOpen
  \bibfield  {author} {\bibinfo {author} {\bibnamefont {{S. Kawamura et
  al.}}},\ }\bibfield  {title} {\bibinfo {title} {{The Japanese space
  gravitational wave antenna: DECIGO}},\ }\href
  {https://doi.org/10.1088/0264-9381/28/9/094011} {\bibfield  {journal}
  {\bibinfo  {journal} {Classical and Quantum Gravity}\ }\textbf {\bibinfo
  {volume} {28}},\ \bibinfo {eid} {094011} (\bibinfo {year}
  {2011})}\BibitemShut {NoStop}%
\bibitem [{\citenamefont {Maggiore}(2018)}]{MaggioreMichele2018GWV2}%
  \BibitemOpen
  \bibfield  {author} {\bibinfo {author} {\bibfnamefont {M.}~\bibnamefont
  {Maggiore}},\ }\href@noop {} {\emph {\bibinfo {title} {Gravitational
  Waves}}},\ Vol.~\bibinfo {volume} {2}\ (\bibinfo  {publisher} {Oxford
  University Press},\ \bibinfo {address} {Oxford},\ \bibinfo {year}
  {2018})\BibitemShut {NoStop}%
\bibitem [{\citenamefont {Romano}\ \emph {et~al.}(2021)\citenamefont {Romano},
  \citenamefont {Hazboun}, \citenamefont {Siemens},\ and\ \citenamefont
  {Archibald}}]{PhysRevD.103.063027}%
  \BibitemOpen
  \bibfield  {author} {\bibinfo {author} {\bibfnamefont {J.~D.}\ \bibnamefont
  {Romano}}, \bibinfo {author} {\bibfnamefont {J.~S.}\ \bibnamefont {Hazboun}},
  \bibinfo {author} {\bibfnamefont {X.}~\bibnamefont {Siemens}},\ and\ \bibinfo
  {author} {\bibfnamefont {A.~M.}\ \bibnamefont {Archibald}},\ }\bibfield
  {title} {\bibinfo {title} {Common-spectrum process versus cross-correlation
  for gravitational-wave searches using pulsar timing arrays},\ }\href
  {https://doi.org/10.1103/PhysRevD.103.063027} {\bibfield  {journal} {\bibinfo
   {journal} {Phys. Rev. D}\ }\textbf {\bibinfo {volume} {103}},\ \bibinfo
  {pages} {063027} (\bibinfo {year} {2021})}\BibitemShut {NoStop}%
\bibitem [{\citenamefont {{Hils}}\ \emph {et~al.}(1990)\citenamefont {{Hils}},
  \citenamefont {{Bender}},\ and\ \citenamefont
  {{Webbink}}}]{1990ApJ...360...75H}%
  \BibitemOpen
  \bibfield  {author} {\bibinfo {author} {\bibfnamefont {D.}~\bibnamefont
  {{Hils}}}, \bibinfo {author} {\bibfnamefont {P.~L.}\ \bibnamefont
  {{Bender}}},\ and\ \bibinfo {author} {\bibfnamefont {R.~F.}\ \bibnamefont
  {{Webbink}}},\ }\bibfield  {title} {\bibinfo {title} {{Gravitational
  Radiation from the Galaxy}},\ }\href {https://doi.org/10.1086/169098}
  {\bibfield  {journal} {\bibinfo  {journal} {\apj}\ }\textbf {\bibinfo
  {volume} {360}},\ \bibinfo {pages} {75} (\bibinfo {year} {1990})}\BibitemShut
  {NoStop}%
\bibitem [{\citenamefont {{Ruiter}}\ \emph {et~al.}(2009)\citenamefont
  {{Ruiter}}, \citenamefont {{Belczynski}}, \citenamefont {{Benacquista}},\
  and\ \citenamefont {{Holley-Bockelmann}}}]{2009ApJ...693..383R}%
  \BibitemOpen
  \bibfield  {author} {\bibinfo {author} {\bibfnamefont {A.~J.}\ \bibnamefont
  {{Ruiter}}}, \bibinfo {author} {\bibfnamefont {K.}~\bibnamefont
  {{Belczynski}}}, \bibinfo {author} {\bibfnamefont {M.}~\bibnamefont
  {{Benacquista}}},\ and\ \bibinfo {author} {\bibfnamefont {K.}~\bibnamefont
  {{Holley-Bockelmann}}},\ }\bibfield  {title} {\bibinfo {title} {{The
  Contribution of Halo White Dwarf Binaries to the Laser Interferometer Space
  Antenna Signal}},\ }\href {https://doi.org/10.1088/0004-637X/693/1/383}
  {\bibfield  {journal} {\bibinfo  {journal} {\apj}\ }\textbf {\bibinfo
  {volume} {693}},\ \bibinfo {pages} {383} (\bibinfo {year}
  {2009})}\BibitemShut {NoStop}%
\bibitem [{\citenamefont {{Ruiter}}\ \emph {et~al.}(2010)\citenamefont
  {{Ruiter}}, \citenamefont {{Belczynski}}, \citenamefont {{Benacquista}},
  \citenamefont {{Larson}},\ and\ \citenamefont
  {{Williams}}}]{2010ApJ...717.1006R}%
  \BibitemOpen
  \bibfield  {author} {\bibinfo {author} {\bibfnamefont {A.~J.}\ \bibnamefont
  {{Ruiter}}}, \bibinfo {author} {\bibfnamefont {K.}~\bibnamefont
  {{Belczynski}}}, \bibinfo {author} {\bibfnamefont {M.}~\bibnamefont
  {{Benacquista}}}, \bibinfo {author} {\bibfnamefont {S.~L.}\ \bibnamefont
  {{Larson}}},\ and\ \bibinfo {author} {\bibfnamefont {G.}~\bibnamefont
  {{Williams}}},\ }\bibfield  {title} {\bibinfo {title} {{The LISA
  Gravitational Wave Foreground: A Study of Double White Dwarfs}},\ }\href
  {https://doi.org/10.1088/0004-637X/717/2/1006} {\bibfield  {journal}
  {\bibinfo  {journal} {\apj}\ }\textbf {\bibinfo {volume} {717}},\ \bibinfo
  {pages} {1006} (\bibinfo {year} {2010})},\ \Eprint
  {https://arxiv.org/abs/0705.3272} {arXiv:0705.3272 [astro-ph]} \BibitemShut
  {NoStop}%
\bibitem [{\citenamefont {{Toonen}}\ \emph {et~al.}(2012)\citenamefont
  {{Toonen}}, \citenamefont {{Nelemans}},\ and\ \citenamefont {{Portegies
  Zwart}}}]{2012A&A...546A..70T}%
  \BibitemOpen
  \bibfield  {author} {\bibinfo {author} {\bibfnamefont {S.}~\bibnamefont
  {{Toonen}}}, \bibinfo {author} {\bibfnamefont {G.}~\bibnamefont
  {{Nelemans}}},\ and\ \bibinfo {author} {\bibfnamefont {S.}~\bibnamefont
  {{Portegies Zwart}}},\ }\bibfield  {title} {\bibinfo {title} {{Supernova Type
  Ia progenitors from merging double white dwarfs. Using a new population
  synthesis model}},\ }\href {https://doi.org/10.1051/0004-6361/201218966}
  {\bibfield  {journal} {\bibinfo  {journal} {Astronomy \& Astrophysics}\
  }\textbf {\bibinfo {volume} {546}},\ \bibinfo {eid} {A70} (\bibinfo {year}
  {2012})},\ \Eprint {https://arxiv.org/abs/1208.6446} {arXiv:1208.6446
  [astro-ph.HE]} \BibitemShut {NoStop}%
\bibitem [{\citenamefont {{Lamberts}}\ \emph {et~al.}(2019)\citenamefont
  {{Lamberts}}, \citenamefont {{Blunt}}, \citenamefont {{Littenberg}},
  \citenamefont {{Garrison-Kimmel}}, \citenamefont {{Kupfer}},\ and\
  \citenamefont {{Sanderson}}}]{2019MNRAS.490.5888L}%
  \BibitemOpen
  \bibfield  {author} {\bibinfo {author} {\bibfnamefont {A.}~\bibnamefont
  {{Lamberts}}}, \bibinfo {author} {\bibfnamefont {S.}~\bibnamefont {{Blunt}}},
  \bibinfo {author} {\bibfnamefont {T.~B.}\ \bibnamefont {{Littenberg}}},
  \bibinfo {author} {\bibfnamefont {S.}~\bibnamefont {{Garrison-Kimmel}}},
  \bibinfo {author} {\bibfnamefont {T.}~\bibnamefont {{Kupfer}}},\ and\
  \bibinfo {author} {\bibfnamefont {R.~E.}\ \bibnamefont {{Sanderson}}},\
  }\bibfield  {title} {\bibinfo {title} {{Predicting the LISA white dwarf
  binary population in the Milky Way with cosmological simulations}},\ }\href
  {https://doi.org/10.1093/mnras/stz2834} {\bibfield  {journal} {\bibinfo
  {journal} {Monthly Notices of the Royal Astronomical Society}\ }\textbf
  {\bibinfo {volume} {490}},\ \bibinfo {pages} {5888} (\bibinfo {year}
  {2019})},\ \Eprint {https://arxiv.org/abs/1907.00014} {arXiv:1907.00014
  [astro-ph.HE]} \BibitemShut {NoStop}%
\bibitem [{\citenamefont {{Korol}}\ \emph {et~al.}(2022)\citenamefont
  {{Korol}}, \citenamefont {{Hallakoun}}, \citenamefont {{Toonen}},\ and\
  \citenamefont {{Karnesis}}}]{2022MNRAS.511.5936K}%
  \BibitemOpen
  \bibfield  {author} {\bibinfo {author} {\bibfnamefont {V.}~\bibnamefont
  {{Korol}}}, \bibinfo {author} {\bibfnamefont {N.}~\bibnamefont
  {{Hallakoun}}}, \bibinfo {author} {\bibfnamefont {S.}~\bibnamefont
  {{Toonen}}},\ and\ \bibinfo {author} {\bibfnamefont {N.}~\bibnamefont
  {{Karnesis}}},\ }\bibfield  {title} {\bibinfo {title} {{Observationally
  driven Galactic double white dwarf population for LISA}},\ }\href
  {https://doi.org/10.1093/mnras/stac415} {\bibfield  {journal} {\bibinfo
  {journal} {Monthly Notices of the Royal Astronomical Society}\ }\textbf
  {\bibinfo {volume} {511}},\ \bibinfo {pages} {5936} (\bibinfo {year}
  {2022})},\ \Eprint {https://arxiv.org/abs/2109.10972} {arXiv:2109.10972
  [astro-ph.HE]} \BibitemShut {NoStop}%
\bibitem [{\citenamefont {{Robson}}\ \emph {et~al.}(2019)\citenamefont
  {{Robson}}, \citenamefont {{Cornish}},\ and\ \citenamefont
  {{Liu}}}]{2019CQGra..36j5011R}%
  \BibitemOpen
  \bibfield  {author} {\bibinfo {author} {\bibfnamefont {T.}~\bibnamefont
  {{Robson}}}, \bibinfo {author} {\bibfnamefont {N.~J.}\ \bibnamefont
  {{Cornish}}},\ and\ \bibinfo {author} {\bibfnamefont {C.}~\bibnamefont
  {{Liu}}},\ }\bibfield  {title} {\bibinfo {title} {{The construction and use
  of LISA sensitivity curves}},\ }\href
  {https://doi.org/10.1088/1361-6382/ab1101} {\bibfield  {journal} {\bibinfo
  {journal} {Classical and Quantum Gravity}\ }\textbf {\bibinfo {volume}
  {36}},\ \bibinfo {eid} {105011} (\bibinfo {year} {2019})},\ \Eprint
  {https://arxiv.org/abs/1803.01944} {arXiv:1803.01944 [astro-ph.HE]}
  \BibitemShut {NoStop}%
\bibitem [{\citenamefont {Maggiore}(2008)}]{MaggioreMichele2008GWV1}%
  \BibitemOpen
  \bibfield  {author} {\bibinfo {author} {\bibfnamefont {M.}~\bibnamefont
  {Maggiore}},\ }\href@noop {} {\emph {\bibinfo {title} {Gravitational
  Waves}}},\ Vol.~\bibinfo {volume} {1}\ (\bibinfo  {publisher} {Oxford
  University Press},\ \bibinfo {address} {Oxford},\ \bibinfo {year}
  {2008})\BibitemShut {NoStop}%
\bibitem [{\citenamefont {{K. Aggarwal et al.}}(2019)}]{2019ApJ...880..116A}%
  \BibitemOpen
  \bibfield  {author} {\bibinfo {author} {\bibnamefont {{K. Aggarwal et
  al.}}},\ }\bibfield  {title} {\bibinfo {title} {{The NANOGrav 11 yr Data Set:
  Limits on Gravitational Waves from Individual Supermassive Black Hole
  Binaries}},\ }\href {https://doi.org/10.3847/1538-4357/ab2236} {\bibfield
  {journal} {\bibinfo  {journal} {\apj}\ }\textbf {\bibinfo {volume} {880}},\
  \bibinfo {eid} {116} (\bibinfo {year} {2019})},\ \Eprint
  {https://arxiv.org/abs/1812.11585} {arXiv:1812.11585 [astro-ph.GA]}
  \BibitemShut {NoStop}%
\bibitem [{\citenamefont {{Sesana}}\ \emph {et~al.}(2008)\citenamefont
  {{Sesana}}, \citenamefont {{Vecchio}},\ and\ \citenamefont
  {{Colacino}}}]{2008MNRAS.390..192S}%
  \BibitemOpen
  \bibfield  {author} {\bibinfo {author} {\bibfnamefont {A.}~\bibnamefont
  {{Sesana}}}, \bibinfo {author} {\bibfnamefont {A.}~\bibnamefont
  {{Vecchio}}},\ and\ \bibinfo {author} {\bibfnamefont {C.~N.}\ \bibnamefont
  {{Colacino}}},\ }\bibfield  {title} {\bibinfo {title} {{The stochastic
  gravitational-wave background from massive black hole binary systems:
  implications for observations with Pulsar Timing Arrays}},\ }\href
  {https://doi.org/10.1111/j.1365-2966.2008.13682.x} {\bibfield  {journal}
  {\bibinfo  {journal} {Monthly Notices of the Royal Astronomical Society}\
  }\textbf {\bibinfo {volume} {390}},\ \bibinfo {pages} {192} (\bibinfo {year}
  {2008})},\ \Eprint {https://arxiv.org/abs/0804.4476} {arXiv:0804.4476
  [astro-ph]} \BibitemShut {NoStop}%
\bibitem [{\citenamefont {{Seto}}\ \emph {et~al.}(2001)\citenamefont {{Seto}},
  \citenamefont {{Kawamura}},\ and\ \citenamefont
  {{Nakamura}}}]{2001PhRvL..87v1103S}%
  \BibitemOpen
  \bibfield  {author} {\bibinfo {author} {\bibfnamefont {N.}~\bibnamefont
  {{Seto}}}, \bibinfo {author} {\bibfnamefont {S.}~\bibnamefont {{Kawamura}}},\
  and\ \bibinfo {author} {\bibfnamefont {T.}~\bibnamefont {{Nakamura}}},\
  }\bibfield  {title} {\bibinfo {title} {{Possibility of Direct Measurement of
  the Acceleration of the Universe Using 0.1 Hz Band Laser Interferometer
  Gravitational Wave Antenna in Space}},\ }\href
  {https://doi.org/10.1103/PhysRevLett.87.221103} {\bibfield  {journal}
  {\bibinfo  {journal} {PRL}\ }\textbf {\bibinfo {volume} {87}},\ \bibinfo
  {eid} {221103} (\bibinfo {year} {2001})},\ \Eprint
  {https://arxiv.org/abs/astro-ph/0108011} {arXiv:astro-ph/0108011 [astro-ph]}
  \BibitemShut {NoStop}%
\bibitem [{\citenamefont {{Maggiore}}\ \emph {et~al.}(2020)\citenamefont
  {{Maggiore}}, \citenamefont {{Van Den Broeck}}, \citenamefont {{Bartolo}},
  \citenamefont {{Belgacem}}, \citenamefont {{Bertacca}}, \citenamefont
  {{Bizouard}}, \citenamefont {{Branchesi}}, \citenamefont {{Clesse}},
  \citenamefont {{Foffa}}, \citenamefont {{Garc{\'\i}a-Bellido}}, \citenamefont
  {{Grimm}}, \citenamefont {{Harms}}, \citenamefont {{Hinderer}}, \citenamefont
  {{Matarrese}}, \citenamefont {{Palomba}}, \citenamefont {{Peloso}},
  \citenamefont {{Ricciardone}},\ and\ \citenamefont
  {{Sakellariadou}}}]{2020JCAP...03..050M}%
  \BibitemOpen
  \bibfield  {author} {\bibinfo {author} {\bibfnamefont {M.}~\bibnamefont
  {{Maggiore}}}, \bibinfo {author} {\bibfnamefont {C.}~\bibnamefont {{Van Den
  Broeck}}}, \bibinfo {author} {\bibfnamefont {N.}~\bibnamefont {{Bartolo}}},
  \bibinfo {author} {\bibfnamefont {E.}~\bibnamefont {{Belgacem}}}, \bibinfo
  {author} {\bibfnamefont {D.}~\bibnamefont {{Bertacca}}}, \bibinfo {author}
  {\bibfnamefont {M.~A.}\ \bibnamefont {{Bizouard}}}, \bibinfo {author}
  {\bibfnamefont {M.}~\bibnamefont {{Branchesi}}}, \bibinfo {author}
  {\bibfnamefont {S.}~\bibnamefont {{Clesse}}}, \bibinfo {author}
  {\bibfnamefont {S.}~\bibnamefont {{Foffa}}}, \bibinfo {author} {\bibfnamefont
  {J.}~\bibnamefont {{Garc{\'\i}a-Bellido}}}, \bibinfo {author} {\bibfnamefont
  {S.}~\bibnamefont {{Grimm}}}, \bibinfo {author} {\bibfnamefont
  {J.}~\bibnamefont {{Harms}}}, \bibinfo {author} {\bibfnamefont
  {T.}~\bibnamefont {{Hinderer}}}, \bibinfo {author} {\bibfnamefont
  {S.}~\bibnamefont {{Matarrese}}}, \bibinfo {author} {\bibfnamefont
  {C.}~\bibnamefont {{Palomba}}}, \bibinfo {author} {\bibfnamefont
  {M.}~\bibnamefont {{Peloso}}}, \bibinfo {author} {\bibfnamefont
  {A.}~\bibnamefont {{Ricciardone}}},\ and\ \bibinfo {author} {\bibfnamefont
  {M.}~\bibnamefont {{Sakellariadou}}},\ }\bibfield  {title} {\bibinfo {title}
  {{Science case for the Einstein telescope}},\ }\href
  {https://doi.org/10.1088/1475-7516/2020/03/050} {\bibfield  {journal}
  {\bibinfo  {journal} {Journal of Cosmology and Astroparticle Physics}\
  }\textbf {\bibinfo {volume} {2020}}\bibfield  {number} {\bibinfo  {number} {
  (3)},\ \bibinfo {eid} {050}},\ }\Eprint {https://arxiv.org/abs/1912.02622}
  {arXiv:1912.02622 [astro-ph.CO]} \BibitemShut {NoStop}%
\bibitem [{\citenamefont {{Yagi}}\ and\ \citenamefont
  {{Seto}}(2011)}]{2011PhRvD..83d4011Y}%
  \BibitemOpen
  \bibfield  {author} {\bibinfo {author} {\bibfnamefont {K.}~\bibnamefont
  {{Yagi}}}\ and\ \bibinfo {author} {\bibfnamefont {N.}~\bibnamefont
  {{Seto}}},\ }\bibfield  {title} {\bibinfo {title} {{Detector configuration of
  DECIGO/BBO and identification of cosmological neutron-star binaries}},\
  }\href {https://doi.org/10.1103/PhysRevD.83.044011} {\bibfield  {journal}
  {\bibinfo  {journal} {Physical Review D}\ }\textbf {\bibinfo {volume} {83}},\
  \bibinfo {eid} {044011} (\bibinfo {year} {2011})},\ \Eprint
  {https://arxiv.org/abs/1101.3940} {arXiv:1101.3940 [astro-ph.CO]}
  \BibitemShut {NoStop}%
\bibitem [{\citenamefont {{Isoyama}}\ \emph {et~al.}(2018)\citenamefont
  {{Isoyama}}, \citenamefont {{Nakano}},\ and\ \citenamefont
  {{Nakamura}}}]{2018PTEP.2018g3E01I}%
  \BibitemOpen
  \bibfield  {author} {\bibinfo {author} {\bibfnamefont {S.}~\bibnamefont
  {{Isoyama}}}, \bibinfo {author} {\bibfnamefont {H.}~\bibnamefont
  {{Nakano}}},\ and\ \bibinfo {author} {\bibfnamefont {T.}~\bibnamefont
  {{Nakamura}}},\ }\bibfield  {title} {\bibinfo {title} {{Multiband
  gravitational-wave astronomy: Observing binary inspirals with a decihertz
  detector, B-DECIGO}},\ }\href {https://doi.org/10.1093/ptep/pty078}
  {\bibfield  {journal} {\bibinfo  {journal} {Progress of Theoretical and
  Experimental Physics}\ }\textbf {\bibinfo {volume} {2018}},\ \bibinfo {eid}
  {073E01} (\bibinfo {year} {2018})},\ \Eprint
  {https://arxiv.org/abs/1802.06977} {arXiv:1802.06977 [gr-qc]} \BibitemShut
  {NoStop}%
\bibitem [{\citenamefont {{Stegmann}}\ \emph {et~al.}(2023)\citenamefont
  {{Stegmann}}, \citenamefont {{Zwick}}, \citenamefont {{Vermeulen}},
  \citenamefont {{Antonini}},\ and\ \citenamefont
  {{Mayer}}}]{2023arXiv231106335S}%
  \BibitemOpen
  \bibfield  {author} {\bibinfo {author} {\bibfnamefont {J.}~\bibnamefont
  {{Stegmann}}}, \bibinfo {author} {\bibfnamefont {L.}~\bibnamefont {{Zwick}}},
  \bibinfo {author} {\bibfnamefont {S.~M.}\ \bibnamefont {{Vermeulen}}},
  \bibinfo {author} {\bibfnamefont {F.}~\bibnamefont {{Antonini}}},\ and\
  \bibinfo {author} {\bibfnamefont {L.}~\bibnamefont {{Mayer}}},\ }\bibfield
  {title} {\bibinfo {title} {{Observing supermassive black holes with deci-Hz
  gravitational-wave detectors}},\ }\href
  {https://doi.org/10.48550/arXiv.2311.06335} {\bibfield  {journal} {\bibinfo
  {journal} {arXiv e-prints}\ ,\ \bibinfo {eid} {arXiv:2311.06335}} (\bibinfo
  {year} {2023})},\ \Eprint {https://arxiv.org/abs/2311.06335}
  {arXiv:2311.06335 [astro-ph.HE]} \BibitemShut {NoStop}%
\bibitem [{\citenamefont {{Abbott, B.~P. et al.}}(2017)}]{2017ApJ...848L..12A}%
  \BibitemOpen
  \bibfield  {author} {\bibinfo {author} {\bibnamefont {{Abbott, B.~P. et
  al.}}},\ }\bibfield  {title} {\bibinfo {title} {{Multi-messenger Observations
  of a Binary Neutron Star Merger}},\ }\href
  {https://doi.org/10.3847/2041-8213/aa91c9} {\bibfield  {journal} {\bibinfo
  {journal} {The Astrophysical Journal Letters}\ }\textbf {\bibinfo {volume}
  {848}},\ \bibinfo {eid} {L12} (\bibinfo {year} {2017})},\ \Eprint
  {https://arxiv.org/abs/1710.05833} {arXiv:1710.05833 [astro-ph.HE]}
  \BibitemShut {NoStop}%
\bibitem [{\citenamefont {{Nelemans}}\ \emph
  {et~al.}(2001{\natexlab{b}})\citenamefont {{Nelemans}}, \citenamefont
  {{Portegies Zwart}}, \citenamefont {{Verbunt}},\ and\ \citenamefont
  {{Yungelson}}}]{2001A&A...368..939N}%
  \BibitemOpen
  \bibfield  {author} {\bibinfo {author} {\bibfnamefont {G.}~\bibnamefont
  {{Nelemans}}}, \bibinfo {author} {\bibfnamefont {S.~F.}\ \bibnamefont
  {{Portegies Zwart}}}, \bibinfo {author} {\bibfnamefont {F.}~\bibnamefont
  {{Verbunt}}},\ and\ \bibinfo {author} {\bibfnamefont {L.~R.}\ \bibnamefont
  {{Yungelson}}},\ }\bibfield  {title} {\bibinfo {title} {{Population synthesis
  for double white dwarfs. II. Semi-detached systems: AM CVn stars}},\ }\href
  {https://doi.org/10.1051/0004-6361:20010049} {\bibfield  {journal} {\bibinfo
  {journal} {Astronomy \& Astrophysics}\ }\textbf {\bibinfo {volume} {368}},\
  \bibinfo {pages} {939} (\bibinfo {year} {2001}{\natexlab{b}})},\ \Eprint
  {https://arxiv.org/abs/astro-ph/0101123} {arXiv:astro-ph/0101123 [astro-ph]}
  \BibitemShut {NoStop}%
\bibitem [{\citenamefont {{Nissanke}}\ \emph {et~al.}(2012)\citenamefont
  {{Nissanke}}, \citenamefont {{Vallisneri}}, \citenamefont {{Nelemans}},\ and\
  \citenamefont {{Prince}}}]{2012ApJ...758..131N}%
  \BibitemOpen
  \bibfield  {author} {\bibinfo {author} {\bibfnamefont {S.}~\bibnamefont
  {{Nissanke}}}, \bibinfo {author} {\bibfnamefont {M.}~\bibnamefont
  {{Vallisneri}}}, \bibinfo {author} {\bibfnamefont {G.}~\bibnamefont
  {{Nelemans}}},\ and\ \bibinfo {author} {\bibfnamefont {T.~A.}\ \bibnamefont
  {{Prince}}},\ }\bibfield  {title} {\bibinfo {title} {{Gravitational-wave
  Emission from Compact Galactic Binaries}},\ }\href
  {https://doi.org/10.1088/0004-637X/758/2/131} {\bibfield  {journal} {\bibinfo
   {journal} {\apj}\ }\textbf {\bibinfo {volume} {758}},\ \bibinfo {eid} {131}
  (\bibinfo {year} {2012})},\ \Eprint {https://arxiv.org/abs/1201.4613}
  {arXiv:1201.4613 [astro-ph.GA]} \BibitemShut {NoStop}%
\end{thebibliography}%

\end{document}